\documentclass[trackchanges, twocolumn]{aastex701}

\usepackage{amsmath}

\usepackage{newtxtext,newtxmath}
\usepackage[T1]{fontenc}
\usepackage[normalem]{ulem}
\usepackage{xcolor}
\definecolor{palered}{RGB}{255,200,200}

\usepackage{graphicx}
\usepackage{amsmath}
\usepackage{xspace}
\usepackage{xcolor}
\usepackage{listings}
\usepackage{lipsum}
\usepackage{threeparttable}

\usepackage{comment}
\usepackage[all]{hypcap}

\definecolor{palered}{RGB}{255,200,200}

\usepackage{soul}

\long\def\new#1{{#1}}

\newcommand{\mesa}{\texttt{MESA}}

\begin{document}

\title{Common Envelope Formation of Apparently Single WN Stars at Low Metallicity}

\author[orcid=0000-0001-6251-5315,gname=Natalia,sname=Ivanova]{Natalia Ivanova}
\affiliation{Department of Physics, University of Alberta, Edmonton, T6G 2E7, Alberta, Canada}
\email{nata.ivanova@ualberta.ca}

\author[orcid=0009-0001-6716-0872, gname=Mackay, sname=Henry]{Henry Mackay}
\affiliation{Department of Physics, University of Alberta, Edmonton, T6G 2E7, Alberta, Canada}
\affiliation{Department of Physics \& Astronomy, University of British Columbia, Vancouver, V6T 1Z1, British Columbia, Canada}
\email{hdmackay@ualberta.ca}

\author[orcid=0000-0003-0642-8107, gname=Tomer, sname=Shenar]{Tomer Shenar}
\affiliation{The School of Physics and Astronomy, Tel Aviv University, Tel Aviv 6997801, Israel}
\email{tshenar@tauex.tau.ac.il}

\correspondingauthor{Natalia Ivanova} 

\begin{abstract}
Several Wolf-Rayet (WR) stars in the Small Magellanic Cloud appear single,
despite the expected importance of binary stripping for producing hydrogen
poor stars at low metallicity. We present a new channel for forming single
WR stars through common envelope (CE) interaction with a low mass companion.
We evolve SMC metallicity binaries with $20$ to $45\,M_\odot$ primaries and
$1$ to $5\,M_\odot$ companions through mass transfer and CE evolution. In a
substantial region of parameter space, the companion spirals sufficiently
deep into the primary to release the energy required to eject its hydrogen
rich envelope before merging with the exposed remnant. At the separation
where envelope ejection becomes possible, however, the stripped companion
cannot fit within its Roche lobe and instead merges with the remnant. The
interaction therefore both strips the massive star and eliminates the
surviving binary, leaving a single merger product. This Companion merger
channel occurs under both standard and enthalpy CE energy treatments and is
the dominant outcome under the enthalpy criterion. We evolve representative
merger products and find that those from sufficiently massive primaries
become hot, hydrogen poor, and nitrogen enriched during core helium burning.
Products from primaries of $35\,M_\odot$ and above satisfy our adopted WN
luminosity, temperature, and surface composition criteria for a substantial
fraction of their core helium burning lifetimes. Model atmospheres of
the $35+5\,M_\odot$ product with empirical mass loss show WR spectra. Common envelope interaction
with a low mass companion can therefore create single WN stars through a
previously unexplored formation channel at low metallicity.
\end{abstract}


\section{Introduction} 

\label{sec:introduction}

As proposed by \citet{paczynski_common_1976}, a common envelope (CE) event may
occur when mass transfer in a binary becomes dynamically unstable, forming a
shared envelope around both the donor core and the companion. The companion
then orbits within this envelope, transferring orbital energy and angular
momentum to the gas. Such instability typically arises after a star overfills its Roche lobe,
either because the donor expands rapidly, for example during
post-main-sequence evolution toward the giant or supergiant phase
\citep{paczynski_common_1976, iben_common_1993}, or because the donor radius
and Roche lobe radius respond to mass loss such that the donor increasingly
overfills its lobe \citep{Pacz1965, iben_common_1993}.
The ensuing inspiral may deposit enough orbital energy
to eject the envelope, or it may continue until the companion and core merge.
Whether the binary survives this phase, leaving a stripped core in orbit with
its companion, or instead ends in a merger, is one of the central questions of
this work. 
If the envelope is ejected before coalescence, the stripped donor core is
exposed, but the companion need not remain detached: the interaction may
leave a compact post-CE binary, or the companion may subsequently
merge with the stripped remnant. For a sufficiently evolved massive donor,
the exposed core may be hot, luminous, and helium burning.
Here we explore both outcomes: a stripped core that survives in orbit
with its companion and may develop an optically thick WR wind, and a merger
product formed when the companion does not survive. We ask whether either can
be observed as a Wolf-Rayet star.

Wolf-Rayet stars (WRs), first identified by \citet{1867CRAS...65..292W}, are a spectroscopic class characterized by broad emission lines that are produced in hot, fast, and optically thick stellar winds. These stars are generally evolved, hydrogen poor descendants of massive stars with initial masses of roughly $20\,M_\odot$ or higher at solar metallicity, although the exact initial-mass threshold depends on metallicity, mass loss, rotation, and binary evolution \citep{2007ARAA..45..177C}. Classical WR progenitors begin as massive O-type stars that lose their outer H-rich layers through strong winds and/or binary stripping, exposing helium rich layers and, in many cases, the He-burning core itself \citep{2007ARAA..45..177C}.

WR stars are classified into three main spectral sequences: the nitrogen sequence (WN), the carbon sequence (WC), and the much rarer oxygen sequence (WO). WN stars show material processed by the CNO cycle, whereas WC and WO stars show products of helium burning, especially carbon and oxygen, and are therefore often associated with later evolutionary stages of WR stars \citep{2007ARAA..45..177C, 2015AA...581A.110T}.

There are two commonly discussed formation channels for classical WR stars.
The first, the ``Conti scenario'' \citep{1975MSRSL...9..193C, 2007ARAA..45..177C}, involves a massive star 
losing its own hydrogen envelope through high stellar wind mass loss rates.
The second, which is the case we focus on, involves binary stripping
\citep{1967AcA....17..355P, 1967ZA.....65..251K}. In this channel, the
initially more massive star in a close binary can lose its hydrogen envelope
through binary interaction.
In sufficiently unstable systems, this interaction
may proceed through a common envelope phase.
\new{Binary interaction} can
produce a stripped, helium rich remnant at lower initial masses, and
potentially at lower metallicities, than expected for the single-star
wind channel alone \new{\citep[e.g.][]{10.1093/mnras/staf884}}. However,
whether such a stripped remnant is observed as a WR star also depends on
whether it develops a sufficiently dense, optically thick wind to produce the
broad emission line spectrum required for WR classification
\citep{Shenar_2020}.

More generally, the WR phenomenon is a wind signature rather than a unique
evolutionary marker: the appearance of broad WR emission lines depends on wind
density, mass loss rate, ionisation structure, opacity, and surface
composition \citep{2007ARAA..45..177C, Gotberg2018}. The connection between WR
spectral subtype and evolutionary state is therefore not one-to-one. Different
evolutionary channels may produce objects with similar WR spectra, while some
stripped helium stars may fail to appear as classical WR stars if their winds
are too weak \citep{Gotberg2018, Gotberg2023, Shenar_2020}. Beyond the self stripped and
binary stripped channels, WR-like spectra may also arise from very massive
hydrogen burning stars  \citep{1997ApJ...477..792D,Crowther2010} and from the presence of strong magnetic fields \citep{Shenar2023}, so a WR classification alone does not fix the formation
history.
\new{For the low mass Wolf-Rayet central stars of the carbon sequence
([WC]) in planetary nebulae, \citet{2002PASP..114..602D} proposed that a
planet, brown dwarf, or low mass main sequence star spirals into an
asymptotic giant branch star and is destroyed near its core. In their
scenario, the resulting accretion disc drives shear mixing that
dredges up carbon and oxygen from the intershell.}
\new{For massive stars}, a merger product may reach the surface
composition and mass loss rates of WR stars \citep{Li2024}, as
\new{we explore} in this paper.

Low metallicity WR stars present a particular challenge for these formation
channels. They are observed in the SMC and LMC, where weaker line driven winds
make formation through the single star channel more difficult, especially at
lower initial masses \citep{Schootemeijer_2018, Shenar_2020, Smith_2017}.
\new{The spectra of WR stars in the low metallicity galaxy NGC~6822
have also been analysed \citep{2026arXiv260922918T}.}
Because radiatively driven mass loss rates decrease with metallicity, the
binary channel has been expected to become more important for WR formation at
subsolar metallicity, and the WR binary fraction was therefore expected to
rise toward environments such as the Magellanic Clouds.

This expectation is not clearly borne out by the observations. The measured
WR binary fractions in the SMC, $\approx40\%$, and LMC, $\approx30\%$, are
comparable to the Milky Way value of $\approx40\%$, although the SMC value
rests on a very small sample
\citep{2003MNRAS.338..360F, 2003MNRAS.338.1025F, 2016A&A...591A..22S, 2019A&A...627A.151S}.
The lowest luminosity WR stars in both Magellanic Clouds are among those without
detected companions. The five least luminous SMC WR stars are apparently
single \citep{2016A&A...591A..22S, Schootemeijer_2018}, while a multiyear
RV study of six of the low luminosity WN3/O3 stars in the LMC found no
significant variations \citep{2023ApJ...947...77M}. Recent high sensitivity
radial velocity monitoring of the seven apparently single SMC WR stars has
likewise found no new companions in the probed parameter regime.
\citet{Schootemeijer_2024} find, for each star individually, that a
companion more massive than $\sim5\,M_\odot$ at periods shorter than about
one year is unlikely, with probability below about $5\%$. They further
estimate a very low probability that all seven targets host companions with
periods shorter than 10~yr, a population level statement that is
complementary to, though weaker per star than, the individual mass limits.

Despite the weaker winds, the single star channel may remain viable in some
cases. \citet{2025A&A...703A.243B} show that sufficiently rapid rotation
combined with optically thick winds can strip stars with
$M\gtrsim20\,M_\odot$ at SMC metallicity, although the required rotation
lies toward the rapidly rotating tail of the O star population.

\begin{figure}
    \centering
    \includegraphics[width=0.5\textwidth]{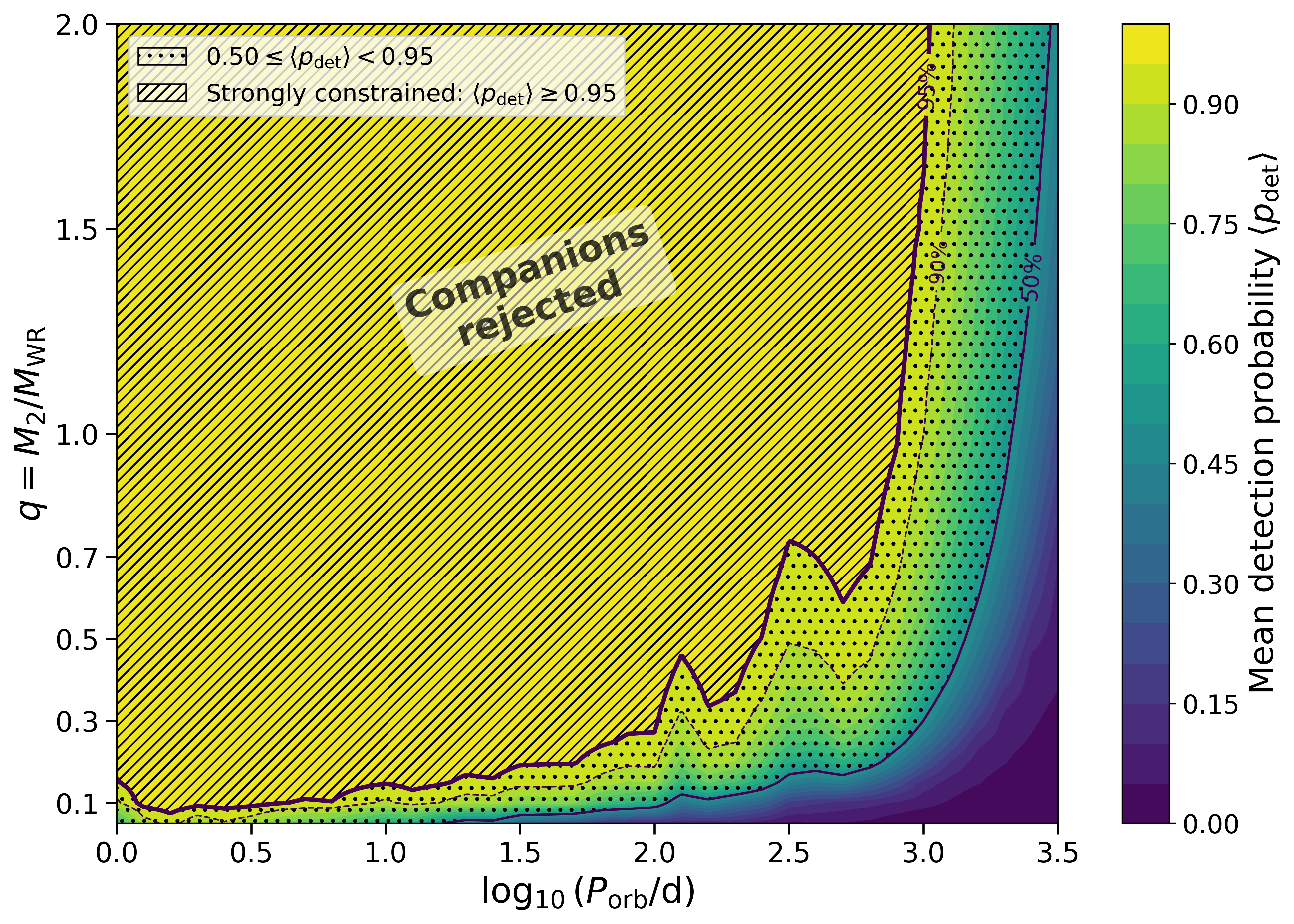}
   \caption{Mean detection probability for companions to the seven apparently single SMC WR stars studied by \citet{Schootemeijer_2024}, shown as a function of orbital period $\log P\,$[d] and mass ratio $q=M_2/M_{\rm WR}$, adopting \(M_{\rm WR}=20\,M_\odot\). We mark regions in which companions are excluded at probability $p > 95\%$ and $p > 50\%$ (see legend). We note that constraints vary from object to object but are qualitatively similar to the mean. Computed using results from \citet{Schootemeijer_2024}.}
    \label{fig:Schoot_Constraints}
\end{figure}

The origin of the apparently single low metallicity WR stars therefore
remains uncertain. The lack of an increased WR binary fraction at low
metallicity does not by itself imply that binary stripping is unimportant,
since the minimum mass for a stripped star to appear as a WR star also
increases toward lower metallicity \citep{Shenar_2020}. Population synthesis
including merger products gives a broad range of single WR fractions
\citep{Li2024}. Several formation paths therefore remain possible.
A star
may strip itself through enhanced mass loss not captured by standard single
star models \new{\citep{Pauli2026, 2026A&A...715L...1M}}.
Alternatively, a companion may strip the
envelope in a common envelope phase and either survive as a faint low mass
star or compact object, or subsequently merge with the exposed remnant
\citep{Schootemeijer_2018, Smith_2017}. This paper addresses the latter two:
the first can leave an extreme mass ratio binary,
$M_2/M_1\lesssim0.2$, while the second leaves a single star, both in
principle consistent with the non-detection of companions around about
$60\%$ of the SMC WR stars \citep{Schootemeijer_2024}.
In the merger case the only direct signature of the interaction is the
transient of the common envelope event itself, which is expected to appear as
a luminous red nova \citep{Tylenda2011, Ivanova2013}.

We therefore calculate common envelope interactions with $1$ to
$5\,M_\odot$ companions and follow both the surviving binaries and selected
merger products to test these two possibilities quantitatively.

Mergers between a main sequence star and an evolved star have been studied
previously \new{\citep[see][for a review]{2026ARA&A..64..629S}}.
 \new{For SN~1987A, \citet{1989ApJ...341..867C} suggested that
a low mass companion spiralling through a common envelope could have
spun up the progenitor's envelope and thereby accounted for the observed
asymmetric expansion of the supernova envelope.} A common envelope merger
origin for the progenitor was proposed by \citet{1989A&A...219L...3H} and
\citet{1990A&A...227L...9P}. \citet{Nomoto1995} extended the idea
to companions of $1$ to $4\,M_\odot$ spiralling into the envelope of an
$11$ to $16\,M_\odot$ star, and proposed that a companion dissolved after
most of the envelope has been ejected leaves a single star with a thin
envelope, the progenitor of a stripped envelope supernova.
\new{\citet{2019SCPMA..6219501S} used the term ``fatal common envelope
evolution'' for events in which a companion merges with the core of a
giant and grouped the resulting supernova progenitors into one class.
He proposed that, following the destruction of a main sequence companion
of about $0.5\,M_\odot$, a wind could remove the hydrogen supplied by
the companion before core collapse, leaving a Type~Ib supernova
progenitor. \citet{2019MNRAS.490....9L} modelled the related Type~IIb
channel for $12$ and $16\,M_\odot$ primaries with a $0.5\,M_\odot$
companion at $Z=0.02$. Their merger products retained $0.02$ to
$0.09\,M_\odot$ of hydrogen at core collapse.}

Detailed merger calculations showed that the entropy of the material
from the dissolved companion determines where it settles, and that this
material can deliver fresh hydrogen into and onto the burning shell of
the primary \citep{2002MNRAS.334..819I,2002PhDT........25I,Ivanova2003}.
The same mechanism was subsequently explored for mergers with more
evolved supergiant primaries, including in detailed calculations of a
merger origin for the progenitor of SN~1987A
\citep{2002PhDT........25I,2017MNRAS.469.4649M}, and its application
to apparently single low metallicity WR stars was proposed, though not
modelled, by \citet{2026ApJ..1002..171B}. Here we connect this merger
outcome to the preceding binary interaction. We map where survival and
merger are obtained across the common envelope parameter space and
follow representative merger products through their subsequent evolution
to assess whether they can enter a WN phase. \new{For one of these products,}
we compute model atmospheres to test whether its emergent spectrum
resembles that of the observed SMC WR stars.

The paper is organised as follows. Section~\ref{sec:methods} describes the model grid, the binary calculations, the criteria used to evaluate the outcomes of the common envelope events, the construction of merger products, and the WR classification criteria. Section~\ref{sec:results} classifies the outcomes across the grid (Section~\ref{sec:results_outcomes}), follows the evolution of four representative merger products (Section~\ref{sec:merger_evolution}), evaluates the luminosities of the merger population against the WN threshold (Section~\ref{sec:population}), compares the surviving binaries with existing radial velocity limits (Section~\ref{sec:survivors}), and compares synthetic spectra of a representative merger product with observed SMC WR stars (Section~\ref{subsec:ModAtm}). Section~\ref{sec:conclusions} summarises the main results.

\section{Methods} \label{sec:methods}

As discussed in Section~\ref{sec:introduction}, the radial velocity limits of
\citet{Schootemeijer_2024} strongly constrain surviving companions around their
apparently single SMC WR targets, whose inferred progenitor masses are
$\gtrsim30$ to $40\,M_\odot$. A low mass companion in this regime implies an
extreme initial mass ratio, making unstable mass transfer and merger especially
relevant. A merger involving the stripped remnant is, moreover, not
constrained by radial velocity limits. We therefore explore primaries from
$20$ to $45\,M_\odot$ with $1$ to $5\,M_\odot$ companions. The upper end of
this range is most directly relevant to the observed SMC WR stars, while the
lower masses map the boundary of the channel and test when the resulting
products cease to satisfy the WN luminosity, temperature, or composition
criteria.

\subsection{Single star evolution}

\label{sec:single_star}

We construct the model grid summarized in Table~\ref{tab}, designed to span
the transition between binary survival and merger for low mass companions
and to cover the primary masses at which the exposed remnant can become
a WR star. 
We adopt $Z = 0.004$, the effective SMC metallicity used by \citet{Schootemeijer_2018}.

The models are computed with Modules for Experiments in Stellar Astrophysics
\citep[MESA;][]{Paxton2011, Paxton2013, Paxton2015, Paxton2018, Paxton2019,
Jermyn2023}. The binary evolution grid through mass transfer and envelope
ejection was computed with MESA version 24.03.1, and the subsequent evolution of the merger products (Section~\ref{sec:merger_evolution}) was computed with MESA version 26.4.1 \footnote{\url{https://doi.org/10.5281/zenodo.2669541}}. The
default microphysics is the same in both releases. The MESA EOS is a blend of
the OPAL \citep{Rogers2002}, SCVH \citep{Saumon1995}, FreeEOS
\citep{Irwin2004}, HELM \citep{Timmes2000}, PC \citep{Potekhin2010}, and Skye
\citep{Jermyn2021} EOSes. Radiative opacities are primarily from OPAL
\citep{Iglesias1993, Iglesias1996}, with low temperature data from
\citet{Ferguson2005} and the high-temperature, Compton-scattering dominated
regime by \citet{Poutanen2017}. Electron conduction opacities are from
\citet{Cassisi2007} and \citet{Blouin2020}. Nuclear reaction rates are from
JINA REACLIB \citep{Cyburt2010}, NACRE \citep{Angulo1999}, and additional
tabulated weak reaction rates \citep{Fuller1985, Oda1994, Langanke2000}.
Screening is included via the prescription of \citet{Chugunov2007}, and thermal
neutrino loss rates are from \citet{Itoh1996}. 

The donors are evolved from zero-age main sequence models with the
\texttt{pp\_and\_cno\_extras} network, extended automatically to
\texttt{co\_burn} and \texttt{approx21} as burning advances, and with Type~2 opacities for $Z_{\rm base} = 0.004$. Convection uses
$\alpha_{\rm MLT} = 1.5$ with the Ledoux criterion, semiconvection with
$\alpha_{\rm sc} = 0.1$, and step overshooting above the convective core
($f = 0.25$, $f_0 = 0.05$). Winds follow the prescription of
\citet{Vink2001} for $T_{\rm eff} > 12\,000$\,K and the \citet{Reimers1975}
and \citet{Blocker1995} prescriptions, scaled by $0.1$ and $0.5$
respectively, below $8\,000$\,K, with a linear blend between these limits.
The atmosphere uses an Eddington $T(\tau)$ relation.
The inlists, customized
routines, and runtime settings used for both sets of calculations will be
made publicly available through Zenodo upon publication.\footnote{Zenodo DOI
to be inserted at acceptance.}

\begin{comment}
Figure~\ref{fig:HR_2x2_grid} shows the HR diagram tracks of the primaries with lines of constant radius overlaid. These tracks provide a reference for the evolutionary state of each donor at the onset of Roche lobe overflow, for each value of $R_L$ in the grid.
\end{comment}

\begin{table}
\centering
\caption{Parameters of the binary model grid.}
\label{tab}
\begin{tabular}{ll}
\hline
Parameter & Values \\
\hline
Primary mass $M_1$ & $20$ to $45\,M_\odot$, in steps of $2.5\,M_\odot$ \\
Companion mass $M_2$ & $1$, $1.5$, $2$, $2.5$, $3$, $3.5$, $4$, $4.5$, $5\,M_\odot$ \\
Seed Roche lobe radius $R_L$ & $20$, $40$, $80$, $160$, $240$, $320$, $400$, \\
 & $480$, $560$, $640$, $800$, $1000$, $1200$, \\
 & $1400$, $1600\,R_\odot$ \\
\hline
\end{tabular}
\end{table}

\subsection{Binary evolution}

After verifying that the isolated primary models are numerically stable, we proceeded with the binary simulations. In these calculations, the initially more massive star is evolved as a full stellar model and the lower-mass companion is treated as a point-mass accretor. The companion masses are between $1$ and $5\,M_\odot$, in steps of $0.5\,M_\odot$ (Table~\ref{tab}). The stellar physics of the donor is that of Section~\ref{sec:single_star}.
Tides are not included, the donor does not rotate, and the orbit loses
angular momentum only through the mass that leaves the system.

To create the initial conditions for each binary run, we first convert the nominal Roche lobe radius $R_L$ of Table~\ref{tab} into an initial orbital separation. This conversion only seeds the separation grid. The actual onset of Roche lobe overflow is determined during the calculation using the Roche equipotential tables of \citet{2023ApJ...952..126P}, which provide the high-precision Roche lobe geometry used by our mass transfer prescription. For the initial estimate, we use the formula of \citet{eggleton_aproximations_1983}:

\begin{equation}
\frac{R_L}{a} = \frac{0.49q^{2/3}}{0.6q^{2/3} + \ln(1+q^{1/3})},
\end{equation}
where $R_L$ is the effective Roche lobe radius, $a$ is the orbital
separation, and $q = M_1/M_2$ is the primary-to-companion mass ratio.
This formula is accurate to better than $1\%$, which is sufficient
for constructing the initial separation grid. At each initial mass
ratio, the seed Roche lobe radii correspond directly to the initial
orbital separations. We space these radii approximately logarithmically, motivated by the commonly adopted distribution of initial orbital separations that is
flat in $\log a$. The actual donor radii at the onset of RLOF
can differ from the nominal grid values, since wind mass loss widens
the orbit before RLOF. We therefore sample the initial orbital
separations rather than the actual radii at RLOF, which are determined
by the subsequent binary evolution and need not follow the same
distribution.

Mass transfer is computed using the Rapid Unified Mass Transfer (RUMT) prescription of \citet{2024ApJ...971...64I}. RUMT treats atmospheric Roche lobe overflow and stream outflow through $L_1$ as a continuous mass transfer process, rather than switching between separate prescriptions. In this approach, the donor structure from the one-dimensional MESA model is mapped onto the binary equipotential geometry, and the mass transfer rate is evaluated across the $L_1$ plane using the local gas properties and the geometry of the corresponding equipotential surfaces. The required three-dimensional equipotential quantities are taken from the tabulations of \citet{2023ApJ...952..126P}. This treatment is crucial for our problem because we follow systems close to the boundary between stable stripping, common envelope-like evolution, and merger, where the degree of overfilling and the geometry of the outflow affect both the mass transfer rate and the subsequent binary survival.
The binary calculations are followed with RUMT from the onset of mass
transfer until one of two conditions indicates dynamical instability: the
donor overflows the outer Lagrangian point, which is the more common case,
or the mass transfer rate rises well above
$10^{-3}\,M_\odot\,{\rm yr}^{-1}$. The donor model at that point defines
the pre-CE structure to which the common envelope treatment of
Section~\ref{sec:outcomes} is applied.
We set the mass transfer efficiency to zero (\texttt{mass\_transfer\_beta} $= 1$ in MESA), so the companion accretes nothing and the transferred mass leaves the system carrying the specific orbital angular momentum of the companion.
\new{Rapid accretion shortly before the common envelope has been suggested
as a possible cause of the inflated radii of main sequence companions to
post-CE central stars of planetary nebulae \citep{2015A&A...580A..19J}.
A companion that accreted and remained inflated would more readily
overfill its Roche lobe after the inspiral.}

\subsection{Evaluating the common envelope outcome}

\label{sec:outcomes}

We assess the outcome of the spiral-in with the energy formalism, in which the envelope can be ejected if the orbital energy released by the inspiral exceeds the binding energy of the overlying envelope \citep{Webbink84, 1990ApJ...358..189D}. For a binary with separation $a$, the orbital energy is $E_{\rm orb} = -G M_1 M_2/(2a)$. If the companion of mass $M_2$ spirals in from the separation $a$ at the
onset of the common envelope, when the donor mass is $M_1$, to a radius $r$
enclosing donor mass $m$, the released orbital energy is
\begin{equation}
\Delta E_{\rm orb}(m) = \frac{G M_{\rm 2}}{2} \left( \frac{m(r)}{r} - \frac{M_{\rm 1}}{a} \right).
\label{eq}
\end{equation}

We define the binding energy of the envelope above mass coordinate $m$ as the positive energy required to remove that material to infinity,
\begin{equation}
E_{\rm bind}(m) = \int_{m}^{M_{\rm 1}} \left( \frac{G m'}{r} - u(m') \right) dm',
\label{eq:ebind}
\end{equation}
where $r$ and $u$ are the radius and specific internal energy at mass coordinate $m'$.
\new{We refer to the ejection criterion based on this binding energy
and the widely used orbital energy formalism of \citet{Webbink84} as
the standard criterion.}
We assume that all released orbital energy is available to unbind the
envelope \new{and therefore set} the common envelope efficiency parameter
to unity.
\new{We neglect the energy released
by accretion onto the companion and the possibility that the companion
launches jets, either during the inspiral or in a possible preceding
grazing envelope phase \citep[e.g.][]{2019SCPMA..6219501S}.}

During a slow spiral-in, the remaining envelope can establish
quasi-stationary outflows. The condition to drive such outflows resembles
the Bernoulli equation and carries an additional enthalpy term $P/\rho$
\citep{BKZ67, ivanova_enthalpy_2011}, so that the relevant binding
energy becomes
\begin{equation}
E_{\rm bind}^{\rm h}(m) = \int_{m}^{M_{\rm 1}} \left( \frac{G m'}{r} - u(m') - \frac{P(m')}{\rho(m')} \right) dm'.
\label{eq:ebindh}
\end{equation}
\new{We refer to the ejection criterion based on this binding energy
as the enthalpy criterion.}
Since $P/\rho \ge 0$, the enthalpy binding energy is smaller than
$E_{\rm bind}$, and the envelope can begin to be removed before its
total energy becomes positive. Three-dimensional common envelope
simulations have also obtained recombination-powered outflows
self-consistently during the envelope-expansion phase
\citep{2016MNRAS.462..362I}. In this work, we therefore include the
enthalpy \new{criterion} alongside \new{the standard criterion}.
\new{The additional $P/\rho$ term makes the enthalpy criterion less
restrictive by reducing the orbital energy required for envelope
ejection. Both criteria adopt unit efficiency.} The difference between
the \new{two} binding energies plays the role that the efficiency
parameter plays in other studies.

The energetic criterion alone specifies a range of allowed ejection depths rather than a unique endpoint, so a physically motivated stopping point is also needed. We adopt the maximum compression point $M_{\rm cp}$ introduced by \citet{ivanova_core}, defined as the pre-CE mass coordinate in the hydrogen burning shell at which the sonic velocity, and hence $P/\rho \propto c_{\rm s}^2$, has a local maximum. Its relevance follows from the thermal readjustment of the remnant after envelope removal. If the envelope is stripped down past $M_{\rm cp}$, so that the remnant mass is below $M_{\rm cp}$, the remnant contracts on its thermal timescale and settles into a compact helium core configuration. If instead mass is left above $M_{\rm cp}$, the star re-expands on its thermal timescale and may drive a renewed episode of mass transfer or thermal pulses, removing mass above $M_{\rm cp}$.

The mass above $M_{\rm cp}$ is therefore the minimum mass that must be removed for the spiral-in to terminate. 
The final orbital separation $a_{\rm f}$ follows from the energy balance: it is the separation at which the released orbital energy, Equation~(\ref{eq}), equals the binding energy of the material above $M_{\rm cp}$,
\begin{equation}
a_{\rm f} = M_{\rm cp} \left( \frac{M_1}{a} + \frac{2 E_{\rm bind}(M_{\rm cp})}{G M_2} \right)^{-1} .
\label{eq:af}
\end{equation}
Since Equation~(\ref{eq}) is evaluated with the pre-CE structure, the balance is meaningful only if $a_{\rm f} \ge r(M_{\rm cp})$, the pre-CE radius of the mass coordinate $M_{\rm cp}$. Otherwise the companion would have to spiral inside the material that is to remain, and we take the envelope to be unejectable.

We evaluate this balance twice, once with $E_{\rm bind}$ and once with $E_{\rm bind}^{\rm h}$, and carry both values of $a_{\rm f}$ through the classification below. The enthalpy formalism gives a wider final orbit, so the two bracket the plausible outcomes.
In the energy balance we use the initial companion mass $M_2$. The mass the companion loses during the inspiral, estimated below, would require a somewhat deeper spiral-in to release the same energy, so the values of $a_{\rm f}$ derived here are upper limits.
\new{The energy required to unbind the stripped material from the
companion and carry it out of the binary would also draw on the
orbital energy released during the inspiral. We omit this cost from the
energy balance. Including it would further reduce $a_{\rm f}$.}

For systems in which the orbital energy budget is sufficient to remove the envelope down to $M_{\rm cp}$, we next determine whether the corresponding post-CE configuration is geometrically possible.
We compute the Roche lobe radius of the donor remnant with the \citet{eggleton_aproximations_1983} approximation, using the remnant mass $M_{\rm cp}$ and the companion mass. The remnant radius immediately after envelope removal and during its subsequent thermal readjustment is uncertain. Because we do not follow this evolution explicitly, we assess the possible configuration using two radius prescriptions: the pre-CE radius at the mass coordinate $M_{\rm cp}$ and the thermal-equilibrium radius of a helium star of the same mass, for which we use the zero-age helium main sequence fit of \citet{hurley_comprehensive_2000}. The pre-CE radius at $M_{\rm cp}$ can be smaller or larger than the equilibrium radius, depending on how strongly that layer is compressed under the envelope, so the remnant may either expand or contract during its thermal readjustment.

If the remnant overfills its Roche lobe for both radius prescriptions, no detached binary is possible at the separation required to eject the envelope down to $M_{\rm cp}$. We classify this as a Remnant contact merger. Unlike a Complete merger, the envelope-ejection criterion is satisfied, and it is the geometry of
the corresponding post-CE configuration that prevents binary survival.

If the core survives, we next consider whether the companion itself is
stripped during the inspiral.
During the inspiral the companion moves through the envelope and experiences on its leading face an effective ram pressure
\begin{equation}
P_{\rm ram} = \rho_{\rm a}\,\left(v_{\rm rel}^2 + c_{\rm s,a}^2\right),
\label{eq:pram}
\end{equation}
where the sound speed term accounts for the thermal pressure of the ambient gas, consistent with the Bondi-Hoyle-Lyttleton formalism \citep{1939PCPS...35..405H, 1944MNRAS.104..273B, 2004NewAR..48..843E}. We evaluate $P_{\rm ram}$ at $a_{\rm f}$, where it is largest, with $v_{\rm rel} = \sqrt{G\,[m(a_{\rm f}) + M_2]/a_{\rm f}}$ the orbital velocity of the companion relative to the donor core and
$m(a_{\rm f})$ the donor mass enclosed within $a_{\rm f}$. Neglecting
envelope corotation overestimates the stripping.

The ambient state of the gas is not that of the unperturbed envelope. By
the end of the spiral-in, the injected orbital energy has lifted and
expanded the envelope, lowering the density around the orbit, and
three-dimensional simulations find a reduction of an order of magnitude
or more \citep[e.g.][]{2016MNRAS.462..362I}. We account for this with a fixed factor $f = 10$, so that $\rho_{\rm a} = \rho(a_{\rm f})/f$. The ambient sound speed is scaled consistently with the adiabatic expansion, $c_{\rm s,a} = c_{\rm s}(a_{\rm f})\, f^{-(\Gamma_1-1)/2}$, where $c_{\rm s}(a_{\rm f})$ and the adiabatic exponent $\Gamma_1$ are read from the pre-CE donor profile.

For the companion structure we use main sequence models at the grid
metallicity.
\footnote{The lowest mass companions would not yet have reached the ZAMS
when the most massive donors first undergo Roche lobe overflow. At the age
when a $40\,M_\odot$ donor first fills its Roche lobe, a $1\,M_\odot$
pre-main sequence star is about $25\%$ larger than its main sequence model,
while for $1.5\,M_\odot$ the difference is already within $2\%$. This does
not affect our outcome classification (Section~\ref{sec:results_outcomes}).
All but one of the systems with $1\,M_\odot$ companions are Remnant contact
or Complete mergers under both criteria. The remaining case, and all
Companion mergers with $1.5\,M_\odot$ companions, overfill their Roche
lobes after stripping, so a larger initial companion radius would not
change the outcome.}
The outer layers of the companion are removed where the ram pressure
exceeds the pressure inside the companion. \new{In a three-dimensional
simulation of a head-on collision, \citet{Ballone2023} found ram
pressure stripping of a main sequence star as it passed through the
envelope of an evolved primary. Its outer layers were stripped and
shocked before the star was tidally disrupted near the primary's core.}
We find the stripping point $r^*$ from the condition $P_2(r^*) = P_{\rm ram}$, where $P_2$ is the pressure profile of the unperturbed companion, and the material above $r^*$ is taken to be lost to the outflow. This pressure comparison is a proxy for the true criterion, the ram force against the local gravitational restoring force, but the two track each other in a hydrostatic star, and with only the unperturbed profiles available a more elaborate treatment is not warranted. The stripping is adiabatic, in the sense that the companion retains its entropy profile, since the spiral-in time is much shorter than the thermal timescale of the companion. At the same time the dynamical time of the companion is much shorter than the spiral-in time, so the
companion sheds the shocked material continuously and remains close to
hydrostatic equilibrium. The stripped companion has mass
$M_2^* = m_2(r^*)$ and radius $r^*$.

The final classification compares the stripped companion to its Roche lobe. We evaluate the Roche lobe radius at $a_{\rm f}$ with the post-stripping mass ratio $q^* = M_2^*/M_{\rm cp}$. If $r^* < R_{\rm L}$, the companion survives and the outcome is a close binary composed of the thermally readjusting remnant and the stripped companion, with mass $M_2^*$ and separation $a_{\rm f}$. If instead the stripped companion overfills its Roche lobe, it is disrupted and merges with the remnant, and only the mass $M_2^*$ takes part in the merger, since the material removed during the inspiral has already joined the ejected envelope. The structure and appearance of the merged product are evaluated in the following section.

A surviving companion is not in thermal equilibrium after the rapid
stripping. It retains the entropy stratification of the interior of the
original star, so it is generally undersized for its new mass and is
expected to re-expand as it thermally readjusts. We do not follow this
evolution explicitly. As an order-of-magnitude diagnostic we estimate its
Kelvin-Helmholtz timescale as
\begin{equation}
    t_{\rm KH,2} \sim \frac{G M_2^{*2}}{r^* L_2^*} ,
    \label{eq:tkh2}
\end{equation}
where $L_2^*$ is the luminosity emerging at the mass coordinate $M_2^*$ in
the original companion model, and compare it with the core helium burning
lifetime of a helium star of mass $M_{\rm cp}$, which provides the
characteristic duration of the remnant's helium burning stage. We take the
latter from the fit of \citet{hurley_comprehensive_2000},
\begin{equation}
    \frac{t_{\rm He}}{\rm Myr} = \frac{0.4129
        + 18.81 \left( M_{\rm cp}/M_\odot \right)^{4}
        + 1.853 \left( M_{\rm cp}/M_\odot \right)^{6}}
        {\left( M_{\rm cp}/M_\odot \right)^{6.5}} .
    \label{eq:thems}
\end{equation}
Companions with $t_{\rm KH,2} \gtrsim t_{\rm He}$ are unlikely to complete
their thermal readjustment on this timescale, whereas those with
$t_{\rm KH,2} \ll t_{\rm He}$ have time to readjust and are candidates for
a renewed episode of mass transfer. Whether such a companion in fact
refills its Roche lobe depends on its detailed thermal response and
requires an evolutionary calculation of the stripped star. We therefore use
survival here to mean survival of the dynamical common envelope phase,
rather than guaranteed long-term detachment.
\new{We also neglect the subsequent interaction of a surviving binary
with a circumbinary disc. If some of the envelope remains bound and
forms such a disc, the interaction could further shrink the orbit
\citep{2026arXiv261107326G} and cause some of the binaries we classify
as survivors to merge.}

\begin{figure*}
    \centering
    \includegraphics[width=\textwidth]{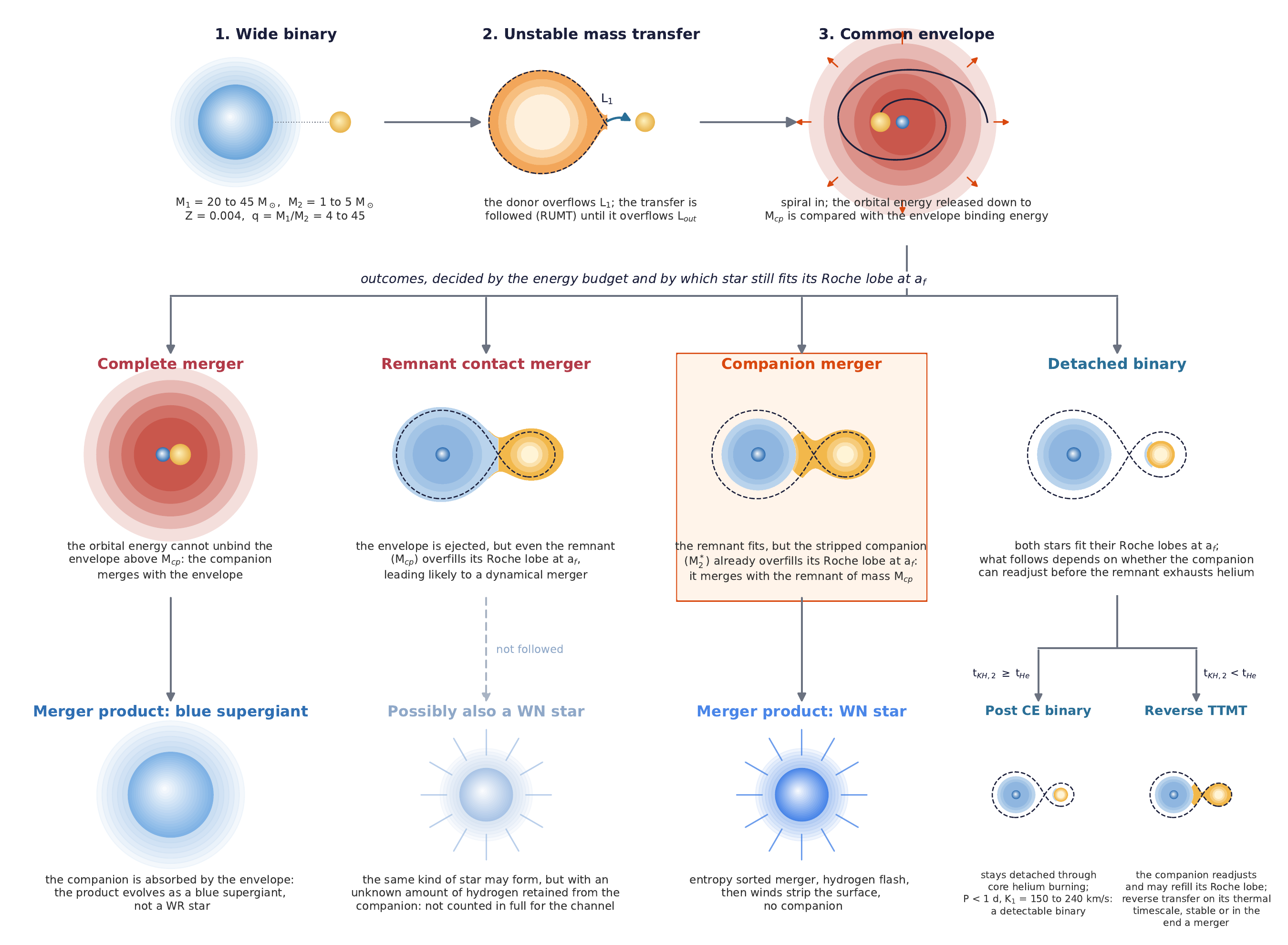}
\caption{The common envelope channel and its outcomes.
Top row: a wide binary with a $20$ to $45\,M_\odot$ primary and a
$1$ to $5\,M_\odot$ companion at $Z=0.004$,
unstable mass transfer, followed until the donor overflows the outer
Lagrangian equipotential, and the common envelope, during which the
orbital energy released by spiralling in to the compression point
$M_{\rm cp}$ is compared with the binding energy of the envelope above it.
Middle row: the four outcomes of the dynamical phase, decided in sequence
by the energy budget and by which star still fits its Roche lobe at the
final separation $a_{\rm f}$ (Section~\ref{sec:outcomes}).
Bottom row: the products that follow.
As in other failed common envelope events, a Complete merger is expected
to retain a substantial hydrogen envelope together with companion
material and evolve as a supergiant, blue if the core to total mass
ratio is reduced sufficiently (Section~\ref{sec:outcomes}).
A remnant contact merger may form the same kind of star, but the amount
of hydrogen retained from the companion is not determined by our
prescription, so this branch (dashed arrow) is not counted in full
for the channel.
Detached binaries are divided into Post-CE binaries and Reverse TTMT
systems according to whether the stripped companion can thermally
readjust before the remnant exhausts core helium.
In all panels the blue star is the core of the primary with its
hydrogen burning shell, the yellow star is the companion, and dashed
curves are the Roche lobes. A star that overfills its lobe is drawn
on the equipotential it fills.
Sizes are schematic and not to scale.}
    \label{fig_scenario}
\end{figure*}

We therefore divide the outcomes into five classes  (see also Fig~\ref{fig_scenario}):

\begin{itemize}
    \item \textbf{Post-CE binary, limited thermal readjustment ("Post-CE binary"):}
    both stars survive the dynamical common envelope phase and
    $t_{\rm KH,2} \ge t_{\rm He}$, so the companion is not expected to
    complete its thermal readjustment during the helium burning stage of
    the remnant.

    \item \textbf{Post-CE binary with potential reverse thermal timescale mass transfer (``Reverse TTMT''):} both stars survive the dynamical CE phase and remain detached at $a_{\rm f}$. However, $t_{\rm KH,2} < t_{\rm He}$, so the companion  has time to thermally readjust before the donor remnant completes core  helium burning and may subsequently refill its Roche lobe. If this occurs, the resulting reverse mass transfer from the much less massive companion onto the remnant is expected to remain stable and proceed on the thermal timescale of the companion rather than lead to a dynamical merger.

    \item \textbf{Merger of the stripped companion with the remnant (``Companion merger''):} the envelope above $M_{\rm cp}$ is ejected and the donor remnant can remain within its Roche lobe, but the stripped companion already overfills its Roche lobe at $a_{\rm f}$. It therefore cannot survive the dynamical CE phase as a detached star and merges with the remnant of mass $M_{\rm cp}$. Only the retained companion mass $M_2^*$ participates in the merger. The structure and appearance of the merged product are evaluated in Section~\ref{sec:merger_evolution}.

    \item \textbf{Remnant contact merger:} the orbital energy budget is sufficient to remove the envelope down to $M_{\rm cp}$, but at the corresponding separation $a_{\rm f}$ the donor remnant cannot fit within its Roche lobe under either of the two radius prescriptions. A detached post-CE binary therefore cannot be established, and the system is treated as a merger.

    \item \textbf{Complete merger:}
    the orbital energy budget is insufficient to eject the envelope down
    to $M_{\rm cp}$, that is, the orbital energy released by spiralling in to $r(M_{\rm cp})$ is smaller than $E_{\rm bind}(M_{\rm cp})$ and Equation~(\ref{eq:af}) has no solution with $a_{\rm f} \ge r(M_{\rm cp})$, so no stripped remnant is produced.
    The product retains the hydrogen envelope together with the
companion material, less the fraction of the envelope ejected during
the spiral in and coalescence. Some envelope ejection is expected
during this process, since the orbital energy released before
coalescence is deposited in the envelope
\citep{Nomoto1995}. However, our prescription does not
estimate the ejected fraction, and the Complete merger class should
not be read as full retention of the envelope.
For core to total mass ratios below about $0.6$, as for all donors
in our grid, a larger envelope mass at a given helium core mass
makes a post-merger star hotter and favours a blue over a red
supergiant \citep{Farrell2022}. Merger models of this kind were shown to reproduce the blue supergiant progenitor of SN~1987A by \citet{2002PhDT........25I} and subsequently by
\citet{2017MNRAS.469.4649M}.

\end{itemize}

For the detached binaries, the division between Post-CE binary and Reverse TTMT at  $t_{\rm KH,2} = t_{\rm He}$ is nominal, since both timescales are order-of-magnitude estimates.

The reverse mass transfer that begins when the stripped companion overfills its Roche lobe at $a_{\rm f}$ cannot stabilise. This is fundamentally different from the Reverse TTMT class, in which the binary emerges from the common envelope detached and the companion subsequently refills its Roche lobe through thermal expansion. Here the companion already overfills its lobe by a finite amount at the end of the dynamical plunge, while the orbit is still decaying in the residual envelope. The mass transfer rate is therefore not regulated by the slow radius evolution of the donor, but by the degree of overfilling, with mass outside the $L_1$ equipotential removed on approximately the dynamical timescale. 

The remnant, itself thermally readjusting and in many systems close to filling its own Roche lobe, cannot accrete at this rate. The transferred material fills the remnant's Roche lobe, bringing the system into contact and driving mass loss through the outer Lagrangian point $L_2$. Although the mass ratio $q^*<1$ would favour orbital expansion under conservative mass transfer, this requires the material lost by the lighter companion to be retained by the heavier remnant. Instead, material escaping through $L_2$, which lies beyond the orbit of the companion, carries away substantially more specific angular momentum than the mean specific orbital angular momentum of the binary. \citet{Lu2023} find that when a large fraction of the transferred mass leaves through $L_2$, the orbit shrinks for practically all mass ratios, including those for which conservative transfer would widen it. Although their model considers quasi-steady transfer through an accretion disc, the underlying mechanism is the removal of orbital angular momentum by the $L_2$ outflow, which also operates when the overflow is dynamical and no disc forms. 

The decreasing separation shrinks the Roche lobe of the companion, increases the degree of overfilling, and further raises the mass loss rate. This positive feedback drives a runaway that cannot be arrested by thermal contraction of the companion, since $t_{\rm KH,2}$ greatly exceeds the duration of the overflow. The stripped companion therefore cannot establish stable reverse mass transfer and instead merges with the remnant.

The classification separates outcomes that an energy comparison
alone would not. \citet{Nomoto1995} assumed two outcomes for a
low mass companion: if the released orbital energy exceeds the envelope
binding energy, the envelope is ejected before the companion is dissolved
and the binary survives as a helium star with a close main sequence
companion; otherwise the companion is dissolved inside the envelope and
the merged star retains the part of the envelope that was not ejected,
which corresponds to our Complete merger class. The merger outcomes can be viewed as a continuous progression in envelope
removal. A Complete merger may eject part of the envelope before
coalescence, while a Remnant contact merger has sufficient orbital energy
to eject the envelope down to $M_{\rm cp}$, but no detached remnant can
form and the amount actually ejected is uncertain. In the Companion
merger channel, the envelope above $M_{\rm cp}$ is removed before the
stripped companion merges with the exposed remnant. The Companion merger is therefore the last merger outcome in a
continuous sequence of common envelope outcomes, rather than a
separate process. At a larger $a_{\rm f}$, where the stripped companion
also fits within its Roche lobe, the same sequence ends in a
surviving binary.

We construct and subsequently evolve merger products only for the Companion merger class. In these systems the envelope has already been removed down to $M_{\rm cp}$ and the surviving companion mass $M_2^*$ is defined by the stripping calculation, so the two components entering the merger are specified. For a Remnant contact merger, by contrast, a detached configuration with a remnant of mass $M_{\rm cp}$ is never established, and our prescription does not determine how much material above $M_{\rm cp}$ remains at the onset of coalescence or is subsequently ejected.

\subsection{Post-CE properties}

The classification above leaves two outcomes to follow further: a surviving close binary, and a merger of the stripped companion with the remnant. We describe the treatment of each in turn; the results are presented in the next section.

\subsubsection{Merger products}

\label{sec:merger_products}

When the stripped companion overfills its Roche lobe at $a_{\rm f}$, it is disrupted and its mass $M_2^*$ is deposited onto the remnant of mass $M_{\rm cp}$. This material cannot settle onto the remnant with its entropy unchanged. Hydrodynamical simulations of the stream-core interaction during slow mergers of massive stars show that the incoming material is shock
heated as it is decelerated and accreted \citep{2002MNRAS.334..819I}. We account for this by raising the specific entropy of the companion material by $\Delta s$. The magnitude follows from the Rankine-Hugoniot conditions:
for the modest Mach numbers of the disruption flow, $\mathcal{M} \approx 2$ to $4$, a single passage through an adiabatic shock in a gas with $\gamma = 5/3$ and about 1.7 particles per baryon gives $\Delta s \approx 0.5$ to $3\,k_{\rm B}$ per baryon. 
We adopt $\Delta s = 1\,k_{\rm B}$ per baryon throughout, near the middle of
this range.

The structure of the merger product is then constructed by entropy
sorting. All mass shells of the remnant and of the heated companion
material are ordered by specific entropy, with the lowest entropy at the
centre \citep{1996ApJ...468..797L, 2002ApJ...568..939L,
2008MNRAS.383L...5G}. The latter two studies sort by the buoyancy
variable $A = P/\rho^{\Gamma}$, which accounts for the composition
dependence of the density at a given pressure. The two orderings coincide
for material of uniform composition, and for the hydrogen rich companion
material sorting by entropy places it somewhat deeper than sorting by
buoyancy would. At the level of this proof of concept the difference is
not decisive, and the actual depth to which the companion material
penetrates is in any case a question for three-dimensional merger
simulations. Where the companion material settles is set by its entropy relative to
the remnant.
If that entropy is intermediate between the remnant core and
the envelope, the material comes to rest in the region of the
hydrogen burning shell, delivering fresh hydrogen into hot layers.
\new{This differs from the treatment of \citet{2019MNRAS.490....9L}, who
added the companion material on top of the nearly bare core, where it
formed a new giant envelope.}

What follows depends on what lies beneath the deposited material.
For a remnant in core helium burning it settles on the hydrogen burning
shell above the helium core, and the ensuing hydrogen flash mixes hydrogen
inward (Section~\ref{sec:merger_evolution}). For a remnant that has
already formed a carbon core it settles on the helium burning shell, and
the hydrogen flash can ignite explosive helium shell burning. In the
one-dimensional calculations of \citet{2002PhDT........25I}, based on the
stream core interaction of \citet{2002MNRAS.334..819I}, the shell material
was mixed out into the hydrogen envelope, which was still present. With
the envelope removed down to $M_{\rm cp}$ before the merger, the same event
would remove the helium shell from the carbon core directly. None of
our donors has reached this stage at Roche lobe overflow. The most
evolved donor, the $25\,M_\odot$ star at $1400\,R_\odot$, overflows near
the end of core helium burning (Section~\ref{sec:results_outcomes}),
and we do not construct a merger product from it.
Here, the composition and entropy
profiles are smoothed across the contact between the remnant and companion
material over $0.3\,M_\odot$ before the MESA relaxation,
but the true level of penetration can only be verified with
three-dimensional simulations.

We do not include rotation in the reconstructed merger products. The
fraction of the pre-merger orbital angular momentum retained by the remnant is
uncertain, since angular momentum can be carried away by the ejected material
or redistributed into a surrounding disc during coalescence. Any WN-like
phase found in these calculations is therefore reached without rotational
support.
Retained rotation would modify the subsequent mass loss, but not in a settled
direction. One-dimensional prescriptions enhance the rate near the $\Omega
\Gamma$ limit \citep{Langer1998, MaederMeynet2000}, whereas multidimensional
wind models find that gravity darkening shifts the outflow towards the poles \citep{Owocki1996} and lowers the surface-averaged rate below the spherical value \citep{MullerVink2014}. Rotational
mixing would additionally alter the surface abundances and luminosity. Our
post-merger evolution should therefore be read as a non-rotating calculation.

We construct a hydrostatic model from the sorted profile by relaxing a {\mesa}  model of the merged mass to the sorted composition and entropy profiles, and then follow the subsequent evolution with \mesa. 

The merger products are evolved with the same convective mixing
parameters and wind prescriptions as the donors
(Section~\ref{sec:single_star}), but with the \texttt{mesa\_49} nuclear
network, with thermohaline mixing with coefficient unity, and with a minimum diffusion coefficient of $10^2\,{\rm cm^2\,s^{-1}}$, which gradually smooths the composition discontinuity remaining after the entropy sorting. Superadiabatic layers near the Eddington limit are treated
with the superadiabatic reduction of \citet{Jermyn2023}.  When the luminosity exceeds the Eddington
luminosity, additional mass loss is applied using the super-Eddington wind
of \mesa\ \citep{Paxton2013}, with
$\dot M = (L-L_{\rm Edd})/(GM/R)$. The controls
\texttt{super\_eddington\_scaling\_factor} and
\texttt{super\_eddington\_wind\_Ledd\_factor} are both set to unity.  
A drag term damps runaway surface acceleration in the outermost $10\%$ of the mass.
The atmosphere uses an Eddington $T(\tau)$ relation with fixed opacity and
$P_{\rm extra}$ factor 2.
We perform these calculations for a small number of representative cases as
a proof of concept and use them in Section~\ref{sec:merger_evolution} to
test the simplified luminosity estimates adopted for the full grid.

\subsubsection{Surviving binaries}

For the systems classified as surviving binaries, we compute the two quantities through which the companion could reveal itself: the radial velocity semi-amplitude of the WR star and the light of the companion. 

The circular post-CE orbit gives the orbital period from $a_{\rm f}$, and the semi-amplitude follows as $K_1 = (2\pi G/P_{\rm orb})^{1/3} M_2^* \sin i\,(M_{\rm cp}+M_2^*)^{-2/3}$, which we evaluate at $\sin i = 1$, giving the maximum possible semi-amplitude.

For the companion's own light we bracket the luminosity ratio of the companion to the remnant between the immediate post-CE value, using the interior luminosity of the stripped companion at $m = M_2^*$, and the thermal-equilibrium value for a star of mass $M_2^*$, taken as $L_2 = L_\odot (M_2^*/M_\odot)^4$,
reached only after the Kelvin-Helmholtz time $t_{\rm KH,2}$ of Equation~(\ref{eq:tkh2}). These quantities are compared with the observational constraints in Section~\ref{sec:survivors}.

\subsubsection{WR classification} \label{sec:wr_class}

Each remnant of the Post-CE binary, Reverse TTMT and Companion merger classes is first tested against a luminosity criterion, evaluated at the grid metallicity
$Z = 0.004$. For the luminosity we use two prescriptions.

The first prescription applies the helium star mass-luminosity relation of \citet{2011A&A...535A..56G} to the remnant mass $M_{\rm cp}$ as a core based estimate of the remnant luminosity; $M_{\rm cp}$ comprises the helium core and the hydrogen burning shell, and with this prescription we neglect the luminosity contribution from any hydrogen rich material above $M_{\rm cp}$, that is, the companion material in a merger product.
The relation (Eq.~10 of the original paper, fitted to
chemically homogeneous helium star models of $8$ to $250\,M_\odot$ and
extrapolated below $8\,M_\odot$ where needed) is
\begin{equation}
\log_{10}(L/L_\odot) = F_1 + F_2\log_{10}(M/M_\odot)
+ F_3\left[\log_{10}(M/M_\odot)\right]^2 ,
\label{eq:mlr}
\end{equation}
with $F_1=3.017$, $F_2=2.446$ and $F_3=-0.306$, from Table~A.1 in their
paper, and $M = M_{\rm cp}$.

The second prescription is based on the mass-luminosity relations of
\citet{2025A&A...703A..40S}, from which we adopt the luminosity of the fully
stripped, pure helium configuration and the maximum luminosity attainable
when a hydrogen burning shell sits on top of the helium core. Hydrogen 
retaining configurations are directly relevant to the observed population:
the apparently single SMC WN stars have surface hydrogen abundances
$X_{\rm H} \simeq 0.2$ to $0.55$ \citep{2015A&A...581A..21H}. This motivates
accounting for the luminosity contribution associated with a residual
hydrogen rich layer rather than treating the remnant solely as a bare helium
star. Their fit has the form
\begin{equation}
\begin{split}
\log_{10}(L/L_\odot) = {} & F_1 + F_2\,x + F_3\,x^2
 + \left(F_4 + F_5\,x + F_6\,x^2\right) X_{\rm H} \\
 & + \left(F_7 + F_8\,x + F_9\,x^2\right) e^{-X_{\rm H}/F_{10}} ,
\end{split}
\label{eq:s25}
\end{equation}
with $x = \log_{10}(M_{\rm tot}/M_\odot)$ (their Eq.~2). We evaluate it at
the total remnant mass, $M_{\rm tot} = M_{\rm cp} + M_2^*$ for a Companion
merger and $M_{\rm tot} = M_{\rm cp}$ for a surviving remnant, so that the
mass arguments of the two prescriptions differ only for the merger products.
At $X_{\rm H} = 0$ Equation~\ref{eq:s25} reduces to the form of
Equation~\ref{eq:mlr} with $F_1 = 2.434$, $F_2 = 3.382$ and $F_3 = -0.673$,
the effective coefficients of the pure helium relation obtained from the
$Z = 0.004$ minimum-luminosity fit of their Table~C.1, valid for
$1$ to $40\,M_\odot$. We adopt this pure helium luminosity as the lower
estimate for our remnants, rather than the formal minimum-luminosity relation
of \citet{2025A&A...703A..40S}, since the latter corresponds to a chemically
homogeneous hydrogen bearing star, whereas our remnants contain a helium core
and are not chemically homogeneous. 
The upper estimate is their
maximum luminosity relation, Equation~\ref{eq:s25} with the $Z = 0.004$
maximum luminosity coefficients of Table~C.1, valid for $1$ to
$18\,M_\odot$ and $0 \le X_{\rm H} \le 0.7$, which we evaluate at
$X_{\rm H} = 0.35$. About two thirds of our Companion merger products
fall within this mass range. For the remaining third, with masses above
$18\,M_\odot$ and up to about $25\,M_\odot$, we extrapolate the relation.
The adopted value lies toward the hydrogen rich end of the surface
abundances found in the explicitly evolved merger products during their
WN-like phase (Section~\ref{sec:merger_evolution}). The choice is not
critical, since between $X_{\rm H}=0.2$ and $0.5$ the maximum luminosity
changes by less than $0.1$\,dex at a given mass. We use the pure helium and
maximum luminosity relations as lower and upper grid level estimates of the
luminosity of a remnant that retains hydrogen above its helium core. The
accuracy of this adopted range, including the contribution of the
hydrogen burning shell, is tested against the full calculations of
Section~\ref{sec:merger_evolution}.

This luminosity must exceed the metallicity-dependent minimum below which
the winds are expected to be too optically thin for a WR classification
\citep{Aguilera_Dena_2022}:
\begin{equation}
\begin{split}
L_{\rm min,WN} &= 6.85\times10^4\!\left(\frac{Z}{0.02}\right)^{-0.71}\!L_\odot, \\
L_{\rm min,WC} &= 7.51\times10^3\!\left(\frac{Z}{0.02}\right)^{-1.55}\!L_\odot ,
\end{split}
\label{eq:Lmin}
\end{equation}
which at $Z = 0.004$ give $L_{\rm min,WN} = 2.15\times10^5\,L_\odot$ and
$L_{\rm min,WC} = 9.10\times10^4\,L_\odot$.
The WN limit corresponds to the luminosity at which the estimated
wind optical depth reaches the calibrated transition
$\tau_{\rm wind}=1.45$ of \citet{Aguilera_Dena_2022}; the WC limit is
obtained by adopting the same optical-depth threshold with their WC wind
prescription. These limits thus serve as a surrogate for the development
of an optically thick WR wind. The WN limit is calibrated to the
observed minimum luminosities of WN stars in the Milky Way and the
Magellanic Clouds \citep{Shenar_2020}, so at $Z = 0.004$ it falls between
the corresponding LMC and SMC values.
We do not apply the calibrated value $\tau_{\rm wind}=1.45$
directly to the wind properties of our MESA models, since the inferred
optical depth depends on the adopted mass loss and wind velocity
prescriptions, and since the threshold itself is uncertain for WN
stars that retain hydrogen, for which \citet{Pauli_2022} obtain better
agreement with the observed LMC population at $\tau_{\rm wind}\simeq0.5$.
Inverting the luminosity prescriptions at $L_{\rm min,WN}$ gives
$M_{\rm cp} \gtrsim 12.5\,M_\odot$ for the first prescription and, for
the second, $M_{\rm tot} \gtrsim 12.5\,M_\odot$ at the pure helium lower
bound and $M_{\rm tot} \gtrsim 9.0\,M_\odot$ at the maximum luminosity upper
bound.
Since a subtype cannot be assigned without the surface composition,
for the grid level classification we conservatively count as WR
candidates only remnants that satisfy $L > L_{\rm min,WN}$, using the WN
threshold as the more restrictive of the two, and we report the counts for
each luminosity prescription and, for the second, for both
bounds separately.

For the explicitly evolved merger products we additionally examine the surface
composition, which must be compatible with the corresponding spectroscopic
sequence. For WN we adopt a surface nitrogen mass fraction
$X_{\rm N} \geq 2.25\times10^{-3}$ as a conservative observational proxy,
corresponding to the lowest value derived among the putatively single SMC WN
stars studied by \citet{2015A&A...581A..21H}. This value is not a sharp
spectroscopic boundary between WN and non-WN stars, since the WN
classification itself is determined from the emergent spectrum and depends
also on the wind properties. The adopted threshold is consistent with the
expectation of CNO-processed material at SMC metallicity. The summed CNO
abundance of SMC B stars and H\,{\sc ii} regions corresponds to a
CNO-equilibrium nitrogen abundance of
$X_{\rm N} \approx 1.5\times10^{-3}$
\citep[][their Table~1]{2015A&A...581A..21H}, while the nitrogen abundances
derived for the observed WN stars exceed this value by factors of $1.5$ to
$6$. 
Our models are computed at $Z = 0.004$ with the \citet{GS1998} mixture, for which the initial CNO mass fraction is
$X_{\rm CNO} = 2.75\times10^{-3}$, so a fully CNO-processed surface in our
models reaches $X_{\rm N} \approx 2.7\times10^{-3}$, above the adopted
threshold. In contrast, the nominal SMC CNO abundance adopted by
\citet{2015A&A...581A..21H} lies below this threshold. The criterion is
therefore applied as an observational reference value rather than as the CNO
equilibrium abundance of the models.
Material that has passed through the helium burning core also
contains carbon produced by helium burning, and its subsequent CNO
processing can raise the nitrogen abundance above the CNO equilibrium
value of the initial mixture.
For WC, which we include because a merger with a remnant that has
already formed a carbon core can expose carbon rich material through
explosive helium shell burning (Section~\ref{sec:merger_products}),
we adopt as an observational
composition proxy the carbon and oxygen number ratios
$0.1 \leq \mathrm{C/He} \leq 0.35$ and
$\mathrm{O/He} \leq 0.06$ derived for LMC WC4 stars
\citep{2002A&A...392..653C}.
Note that these conventions follow the
respective sources, a mass fraction for nitrogen and number ratios for
carbon and oxygen, and each is applied in its own convention when reading
the model profiles. We omit the rare WO sequence, which is associated with
very advanced evolution after core helium burning 
\citep{2015AA...581A.110T}.

Finally, for the explicitly evolved merger products we require the candidate
WR phase to remain hot, adopting $T_{\rm eff} > 20\,000\,$K as a lower limit.
This cut is intended as a conservative exclusion of cool inflated phases,
rather than as a spectroscopic WR temperature threshold: a luminous cool
supergiant can exceed $L_{\rm min,WN}$ without any resemblance to a WR
spectrum. The adopted limit is deliberately less restrictive than the
temperatures of the observed SMC WN population, the majority of which have
$T_* > 75\,000\,$K \citep{2015A&A...581A..21H}; those values are stellar
temperatures defined at a Rosseland optical depth of 20 in the atmosphere
models and are not directly comparable to the model $T_{\rm eff}$.

In summary, a grid remnant is a WR candidate if its luminosity exceeds
$L_{\rm min,WN}$ of Equation~\ref{eq:Lmin}, and we report this
separately for the first prescription and for the lower and upper bounds of
the second. For the merger products followed through their subsequent
evolution, the candidate WR phase is additionally tested against the surface
composition and effective temperature, allowing WN and WC
classifications to be assessed; the outcomes are given in
Section~\ref{sec:merger_evolution}.

\section{Results} \label{sec:results}

\subsection{Outcomes of common envelope events}
\label{sec:results_outcomes}

\begin{figure*}
    \centering
    \includegraphics[width=\textwidth]{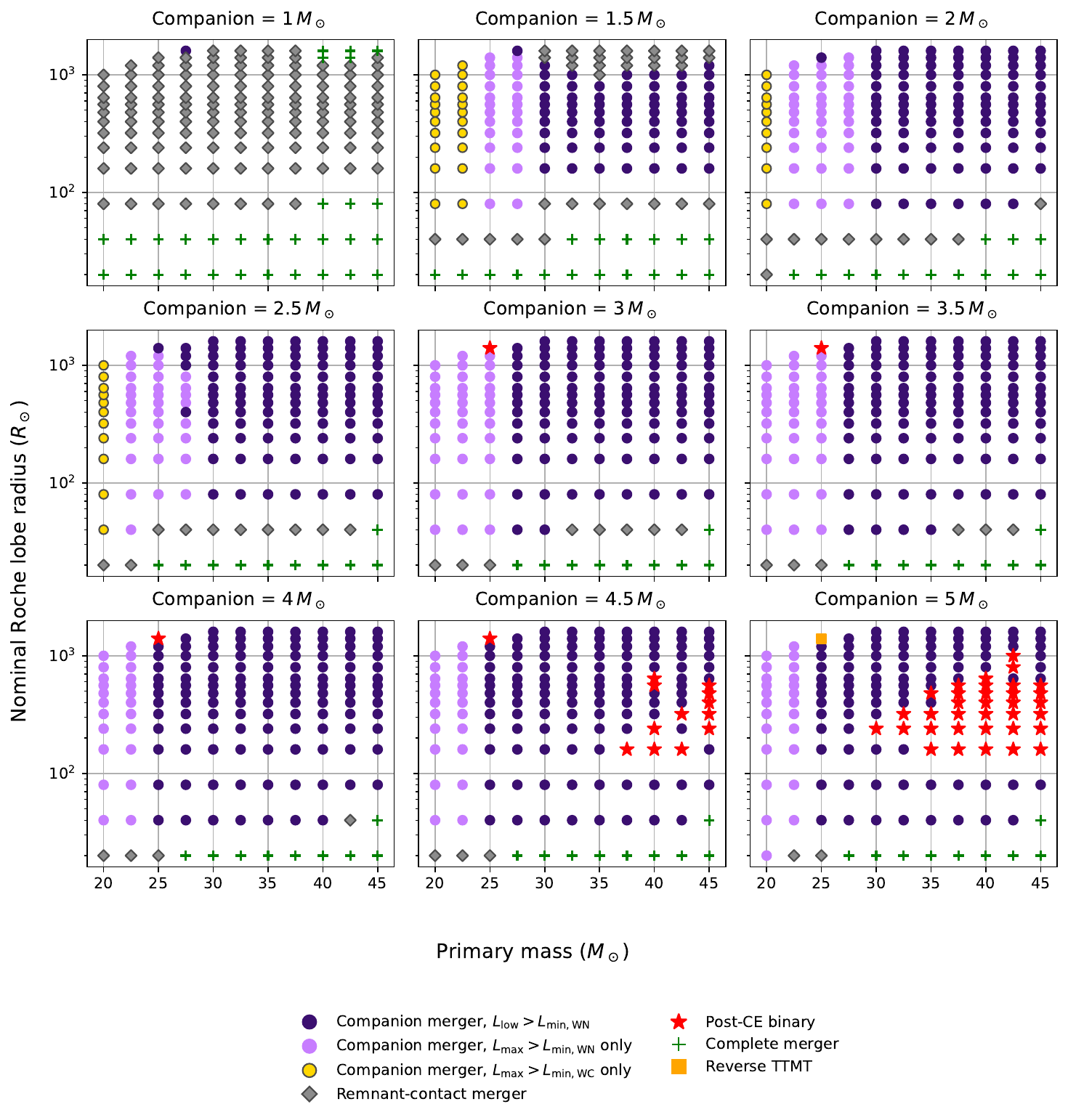}
\caption{Outcomes of the common envelope phase across the model grid under
the enthalpy criterion, shown in nine panels for companion masses from
$1$ to $5\,M_\odot$ in steps of $0.5\,M_\odot$, as a function of primary
mass and nominal Roche lobe radius. Outcome classes are defined in
Section~\ref{sec:outcomes}: Post-CE binary (red star), Reverse TTMT
(orange square), Companion merger (circles), Remnant contact merger (grey
diamond), and Complete merger (green cross). Companion mergers are coloured
by the luminosity estimates of the second prescription of
Section~\ref{sec:wr_class}, evaluated at the merged mass
$M_{\rm tot}=M_{\rm cp}+M_2^*$. Dark purple circles exceed
$L_{\rm min,WN}$ already at the pure helium estimate, light purple circles
exceed it only at the maximum luminosity estimate evaluated at
$X_{\rm H}=0.35$, and yellow circles exceed only the WC threshold,
$L_{\rm min,WC}<L_{\rm max}\leq L_{\rm min,WN}$. No Companion merger falls
below $L_{\rm min,WC}$.}
    \label{fig:parameter_space_outcomes}
\end{figure*}

\begin{figure*}
    \centering
    \includegraphics[width=\textwidth]{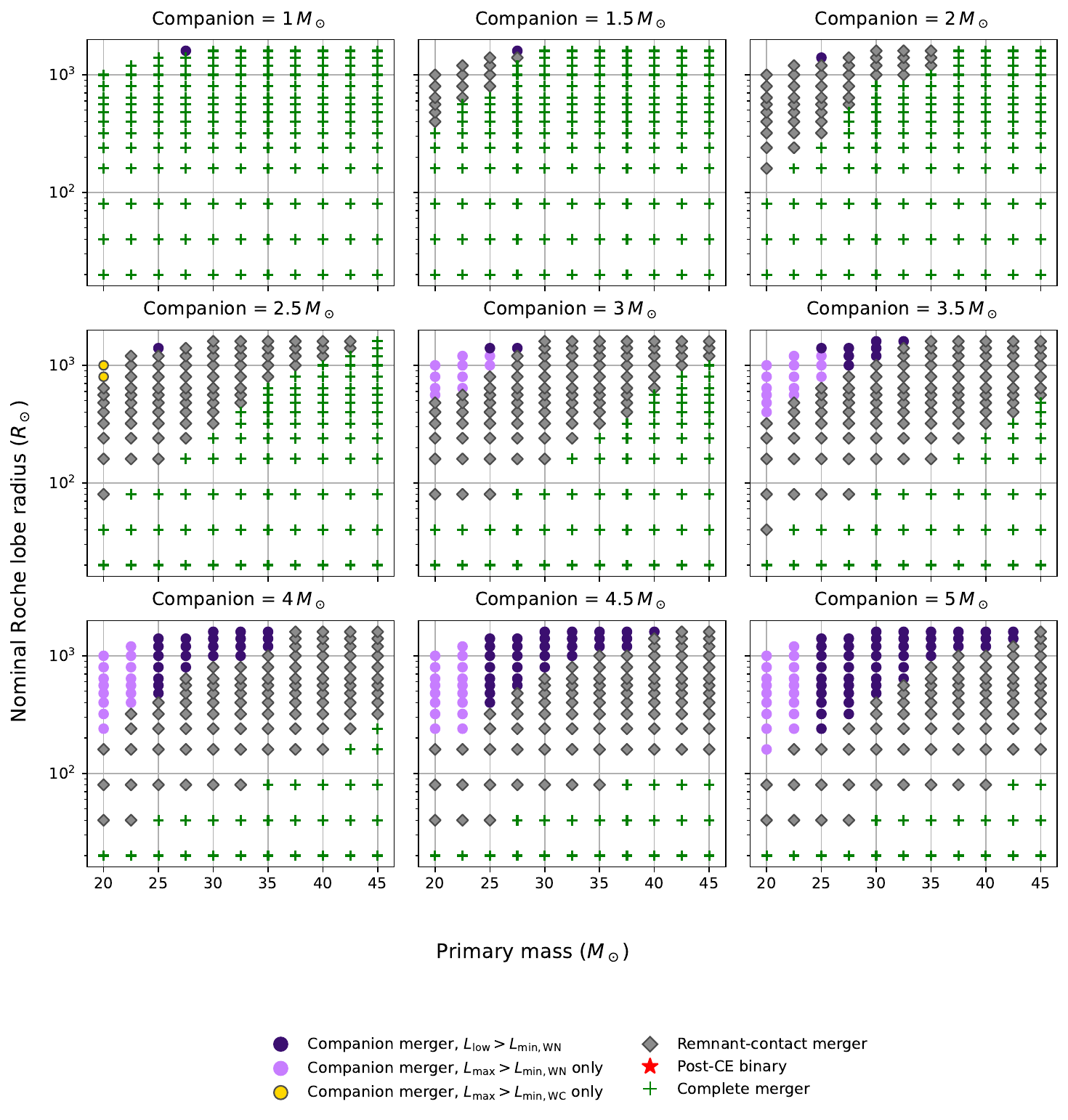}
    \caption{As Figure~\ref{fig:parameter_space_outcomes}, for the standard
    energy criterion. Symbols and the colouring of the Companion mergers
    are the same.}
    \label{fig:parameter_space_outcomes_std}
\end{figure*}

\begin{table}
\centering
\caption{Outcomes of the common envelope phase for the $1424$
grid models confirmed to reach Roche lobe overflow, out of $1485$
possible grid combinations, under the enthalpy and
the standard energy criteria. Percentages refer to the $1424$ models.}
\label{tab:outcomes}
\begin{tabular}{lrrrr}
\hline
Outcome & \multicolumn{2}{c}{Enthalpy} & \multicolumn{2}{c}{Standard} \\
        & $N$ & \% & $N$ & \% \\
\hline
Post-CE binary          &   50 &  4 &   0 &  0 \\
Reverse TTMT            &    1 & $<1$ &   0 &  0 \\
Companion merger        & 1055 & 74 & 187 & 13 \\
Remnant contact merger  &  203 & 14 & 568 & 40 \\
Complete merger         &  115 &  8 & 669 & 47 \\
\hline
\end{tabular}
\end{table}

Table~\ref{tab:outcomes} and Figures~\ref{fig:parameter_space_outcomes} and
\ref{fig:parameter_space_outcomes_std} summarise the outcomes.
Of the $1485$ grid combinations, $34$ never fill the Roche lobe during
the calculation, and for another $27$ Roche lobe overflow could not be
confirmed. This leaves the $1424$ models summarised in
Table~\ref{tab:outcomes}.
Under the
enthalpy criterion, about three quarters of the models are Companion
mergers: the envelope is removed down to $M_{\rm cp}$ and the exposed
remnant can remain within its Roche lobe at $a_{\rm f}$, but the stripped
companion cannot remain detached and subsequently merges with it. Under the
more restrictive standard energy criterion, this channel accounts for about
$13\%$ of the grid. The difference reflects the additional $P/\rho$ term in
the enthalpy binding energy, which makes envelope removal possible with a
shallower inspiral and generally gives a larger $a_{\rm f}$. The energy
condition for removing the envelope down to $M_{\rm cp}$ is satisfied in
about $92\%$ of the models under the enthalpy criterion and about $53\%$
under the standard criterion, but successful removal does not guarantee a
detached post-CE system. 
In about $14\%$ and $40\%$ of the grid,
respectively, the remnant itself overfills its Roche lobe at the
corresponding $a_{\rm f}$ and the outcome is a Remnant contact merger,
while about $8\%$ and $47\%$ fail the envelope ejection condition
altogether and undergo a Complete merger. As in other failed common
envelope events, Complete mergers are expected to retain substantial
hydrogen envelopes and appear as supergiants, blue if the added envelope
mass and the mixing at the merger reduce the core to total mass ratio
sufficiently (Section~\ref{sec:outcomes}), rather than as WR stars.
Among the models that satisfy the energy condition, four in five
are Companion mergers under the enthalpy criterion and one in four
under the standard criterion. These fractions refer to the
calculated grid rather than to a weighted stellar population, but they
demonstrate that the channel proposed in Section~\ref{sec:introduction}, in
which successful envelope removal leaves a stripped remnant with which the
stripped companion then merges, is the main outcome under the enthalpy
criterion and remains a significant outcome under the standard energy
criterion.

The structure of the outcome map is controlled first by whether the energy
balance admits a valid final separation $a_{\rm f}$ and, where it does, by
whether the donor remnant and the stripped companion can each fit within
its Roche lobe at that separation. At fixed donor structure, a more massive
companion allows the required orbital energy to be released at a larger
$a_{\rm f}$, whereas a larger binding energy of the material above
$M_{\rm cp}$ requires a deeper inspiral. The outcomes therefore shift
systematically with increasing companion mass from Complete and
Remnant contact mergers toward Companion mergers and, for the more massive
companions, surviving binaries, as seen across the nine panels of
Figure~\ref{fig:parameter_space_outcomes}. Under the enthalpy criterion,
Remnant contact mergers account for four fifths of the models with
$1\,M_\odot$ companions but only about $1\%$ of those with
$5\,M_\odot$ companions, while the fraction of surviving binaries
increases from zero to about $22\%$. For a $5\,M_\odot$ companion,
the post-stripping radius is about $1.5\,R_\odot$, so survival additionally
requires its Roche lobe at $a_{\rm f}$ to exceed this radius.

The dependence on donor radius is not monotonic. At the smallest radii,
$20$ to $40\,R_\odot$, the material above $M_{\rm cp}$ is tightly bound,
so the energy balance either fails or closes at a very small separation.
For a fixed companion mass, this becomes increasingly restrictive toward
higher primary masses. At intermediate radii the binding energy decreases
and $a_{\rm f}$ reaches its largest values, allowing first the donor remnant
and, under the most favourable conditions, also the companion to remain
within its Roche lobe. At the largest radii $a_{\rm f}$ decreases again.
For the $40+5\,M_\odot$ system, for example, it falls from about
$6\,R_\odot$ at a few hundred $R_\odot$ to below $4\,R_\odot$ at
$1600\,R_\odot$. In our models this decline is associated with an increase
in the binding energy above $M_{\rm cp}$ despite the continuing expansion
of the stellar surface, as the inner structure near the compression point
becomes more compact. Binary survival is therefore confined to a window in
donor radius, even for the most massive companions, while at the largest
radii the outcome can shift back toward Remnant contact or Complete merger
for the lower mass companions. 

The exception among the surviving binaries is the $25\,M_\odot$
donor at $1400\,R_\odot$. With companions of $2\,M_\odot$ and above,
it does not fill its Roche lobe at the first radius maximum after
helium ignition. It then contracts and fills its Roche lobe during
the second expansion about $0.6$\,Myr later, near the end of core
helium burning, when the convective helium core is already receding.
With $1$ and $1.5\,M_\odot$ companions, it overflows during the first
expansion, with the same $M_{\rm cp}$ as at $1200\,R_\odot$.
For the later overflow, the compression point lies about
$1.9\,M_\odot$ further out in mass than at the next smaller grid
radius. Under the enthalpy criterion, $a_{\rm f}$ increases to
$5.5$ to $10\,R_\odot$ for companions of $3$ to $5\,M_\odot$,
and all of these systems survive the dynamical common envelope phase.
The merger cases among the systems that overflow later have little core
helium burning remaining, so we do not construct merger products
from them.

Surviving binaries are rare and occur only under the enthalpy criterion.
In total, $51$ systems survive the common envelope, of which $50$ are
classified as Post-CE binaries and one as Reverse TTMT. Apart from the
exceptional donor discussed above, the surviving systems have primaries
of $30\,M_\odot$ and above, companions of $4.5$ to $5\,M_\odot$, and
Roche lobe overflow at radii of $160$ to $1000\,R_\odot$.
Their final separations are $a_{\rm f}=5.1$ to $6.4\,R_\odot$.
The exceptional donor extends the surviving companion masses down
to $3\,M_\odot$ and the final separations up to about $10\,R_\odot$.
Across the full sample, the companions retain more than $99\%$ of
their initial mass. For the survivors, the estimated thermal readjustment
time of the companion exceeds the adopted helium burning timescale of
the remnant by a factor of $1.3$ to $2.2$, excluding the exceptional
donor. For this donor the ratio is $0.98$ to $3.0$, and the
$25\,M_\odot$ system with the $5\,M_\odot$ companion falls just below
the division adopted in Section~\ref{sec:outcomes} and is classified
as Reverse TTMT. Since this donor is near the end of core helium
burning, the adopted $t_{\rm He}$, which represents the full core
helium burning lifetime of a helium star of mass $M_{\rm cp}$,
overestimates the remaining helium burning lifetime of the remnant.
This classification is therefore nominal.
The observable properties of the
surviving binaries are discussed in Section~\ref{sec:survivors}.

The standard criterion preserves the same qualitative structure of the
outcome map but shifts the boundaries toward less favourable post-CE
outcomes. Without the enthalpy term, the energy condition is harder to
satisfy and, where a valid solution exists, the resulting $a_{\rm f}$ is
typically only $0.8$ to $1.7\,R_\odot$, so that no system satisfies both
geometrical conditions required for binary survival. In most systems for
which the envelope can be removed down to $M_{\rm cp}$, the donor remnant
already overfills its Roche lobe and the outcome is a Remnant contact
merger. Only where the remnant fits within its Roche lobe does the stripped
companion determine whether the outcome is instead a Companion merger. The
fraction of models satisfying the energy condition rises steadily with
companion mass, from essentially none at $1\,M_\odot$ to about half at
$2.5\,M_\odot$ and about $90\%$ at $5\,M_\odot$.

We next follow representative merger products from the Companion merger
channel through their subsequent evolution.

\subsection{Evolution of merger products}
\label{sec:merger_evolution}

We follow the evolution of four representative merger products in full, as
a test of the simplified estimates applied to the grid. The four cases have
a $40\,M_\odot$ donor with a $2\,M_\odot$ companion at a nominal Roche
lobe radius of $400\,R_\odot$, $35\,M_\odot$ and $30\,M_\odot$ donors
with $5\,M_\odot$ companions at $400$ and $640\,R_\odot$, respectively,
and a $20\,M_\odot$ donor with a $2\,M_\odot$ companion at
$400\,R_\odot$. Their remnant masses are $M_{\rm cp}=16.81$, $13.95$,
$11.22$, and $6.19\,M_\odot$, and their merged masses are $18.76$,
$18.93$, $16.20$, and $8.16\,M_\odot$, respectively.

The $35+5\,M_\odot$ calculation illustrates the immediate response to
our merger construction. The relaxed product is initially very compact,
with $R\simeq0.9\,R_\odot$ and $T_{\rm eff}\simeq9000\,{\rm K}$, while
hydrogen rich material from the companion is placed deep in the hot
reconstructed envelope. This hydrogen ignites immediately, producing
$\log L_{\rm H}/L_\odot\simeq12.4$. This is the nuclear luminosity
generated by hydrogen burning in the interior, not the luminosity emerging
from the stellar surface. The envelope subsequently expands and thermally
readjusts in response to this energy release. The hydrogen burning
luminosity remains above $10^{11}\,L_\odot$ for about $7.5$\,days and above
$10^{10}\,L_\odot$ for about $25$\,days. The flash drives convection and
mixes hydrogen inward. The mass coordinate below which $X_{\rm H}<0.01$
moves from $13.6\,M_\odot$ in the initial relaxed model to
$11.0\,M_\odot$ after three days and reaches about $7.6\,M_\odot$ within
the first month. It subsequently moves outward again as hydrogen is
consumed, reaching about $10.0\,M_\odot$ after $5\times10^5$\,yr. The peak
luminosity and detailed timing of this initial flash should not be
overinterpreted, since the construction places the companion material at
its final depth instantaneously. The more relevant consequence for the
subsequent evolution is the redistribution of hydrogen and the associated
thermal readjustment. 
The flash itself is interior and is not observable. The envelope
ejection and the merger that precede it would instead be seen as a luminous
red nova with an ejected mass of about $M_1-M_{\rm cp}$, comparable to the
most massive red nova progenitors inferred to date
\citep{Smith2016, Blagorodnova2017}.
This redistribution reaches into the former convective helium
burning core of the remnant, below about $9\,M_\odot$, where carbon from
helium burning is abundant. CNO processing of this carbon is what later
raises the surface nitrogen of the $35+5\,M_\odot$ product above the
$2.7\times10^{-3}$ reached by processing the initial mixture alone
(Section~\ref{sec:wr_class}). The depth of the mixing, and hence the
nitrogen enrichment, is set by the flash of the instantaneously placed
companion material and should be verified with three-dimensional
hydrodynamical calculations of the merger.
The other three products also undergo an initial
hydrogen flash, reaching peak hydrogen burning luminosities of
$\log L_{\rm H}/L_\odot\simeq10.1$, $12.6$, and $11.8$ for the $40+2$,
$30+5$, and $20+2\,M_\odot$ cases, respectively.

\begin{figure}
    \centering
    \includegraphics[width=1.0\linewidth]{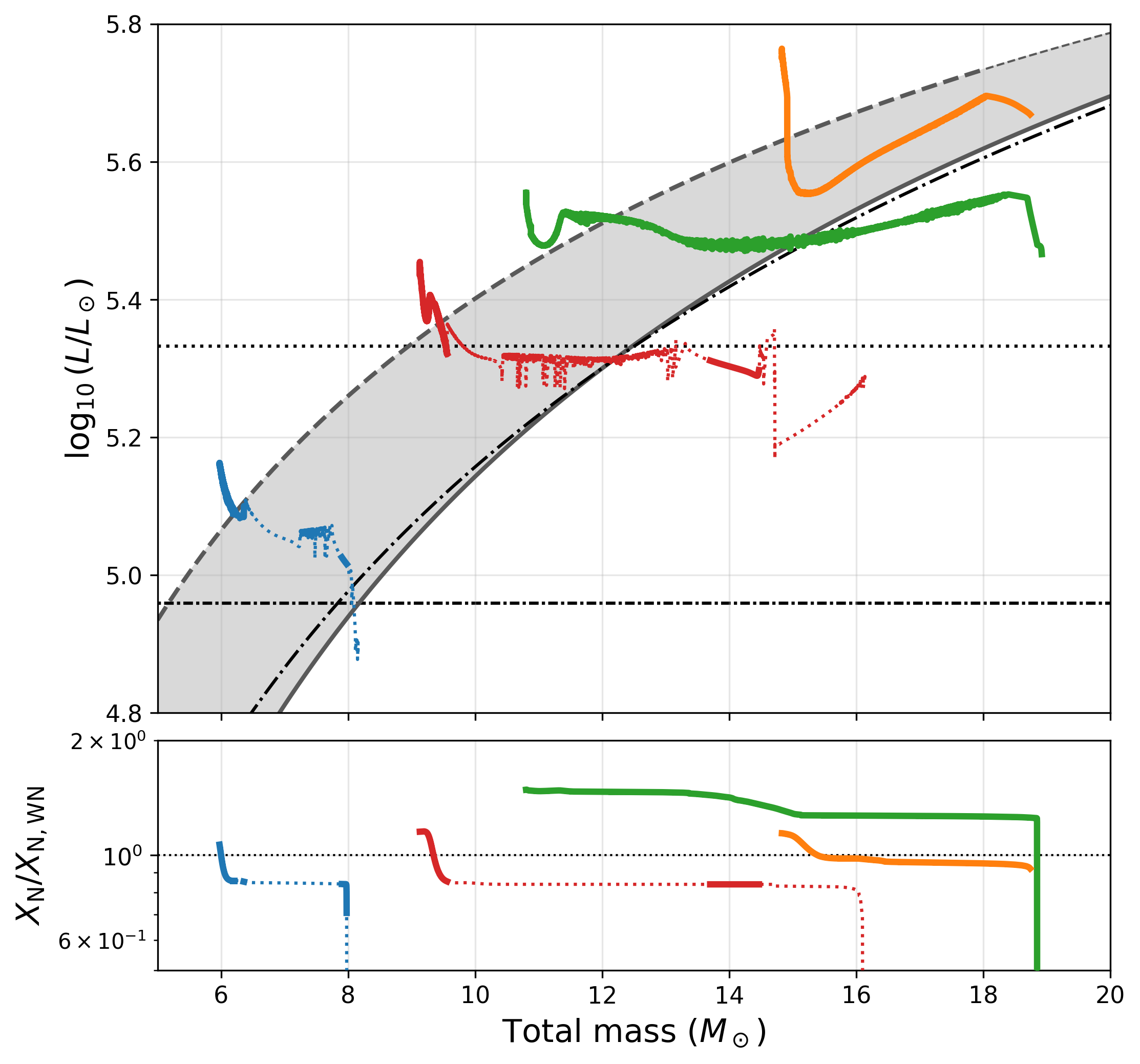}
\caption{
Top: luminosity as a function of mass. The solid grey curve shows the pure
helium estimate of \citet{2025A&A...703A..40S}, and the dashed grey curve
shows their maximum luminosity estimate for $X_{\rm H}=0.35$. The latter is
drawn thin beyond its $18\,M_\odot$ validity limit. The shaded region between
these curves is the luminosity range adopted in Section~\ref{sec:wr_class}.
The dash-dotted black curve shows the helium star relation of
\citet{2011A&A...535A..56G}, used in the first prescription at
$M_{\rm cp}$ and shown here as a function of mass for comparison. Horizontal
lines mark $L_{\rm min,WN}$ (dotted) and $L_{\rm min,WC}$
(dash-dotted) at $Z=0.004$. Coloured tracks show the explicitly evolved
merger products of Section~\ref{sec:merger_evolution}: $40+2\,M_\odot$ at
$R_L=400\,R_\odot$ (orange), $35+5\,M_\odot$ at $R_L=400\,R_\odot$
(green), $30+5\,M_\odot$ at $R_L=640\,R_\odot$ (red), and $20+2\,M_\odot$
at $R_L=400\,R_\odot$ (blue), plotted against their contemporaneous total
mass. Tracks are solid where $T_{\rm eff}>20\,000$\,K and dotted where
they are cooler.
Bottom: surface nitrogen mass fraction along the same evolutionary tracks,
normalised to the adopted WN threshold,
$X_{\rm N,WN}=2.25\times10^{-3}$. The dotted horizontal line marks
$X_{\rm N}=X_{\rm N,WN}$.
}
    \label{fig:luminosityS25}
\end{figure}

The subsequent luminosity and surface nitrogen evolution of all four
products are compared with the adopted WR criteria and luminosity estimates
in Figure~\ref{fig:luminosityS25}.

After this readjustment, the four products follow rather different paths
in the HR diagram. The $40+2$ and $35+5\,M_\odot$ products become hotter
than $20\,000$\,K after about $0.16$\,yr and one year, respectively, and
remain above this temperature for the rest of the calculation. Their radii
never exceed about $18$ and $42\,R_\odot$. The two lower mass products
spend substantial parts of core helium burning at lower effective
temperatures and expand to hundreds of solar radii, reaching about
$903\,R_\odot$ for the $30+5\,M_\odot$ product and $217\,R_\odot$ for the
$20+2\,M_\odot$ product. The $30+5\,M_\odot$ product nevertheless spends
more than half of core helium burning above $20\,000$\,K, while the
$20+2\,M_\odot$ product remains too faint to satisfy the adopted WN
luminosity criterion at any stage. During core helium burning, after the
initial readjustment, the luminosities of the four products span
$\log L/L_\odot=5.55$ to $5.70$, $5.47$ to $5.55$, $5.27$ to $5.33$, and
$4.93$ to $5.07$ for the $40+2$, $35+5$, $30+5$, and $20+2\,M_\odot$
cases, respectively. The faintest apparently single SMC WN stars,
AB 2 and AB 10, have $\log L/L_\odot=5.57$ and $5.65$, respectively
\citep{2015A&A...581A..21H}. The $40+2\,M_\odot$ product therefore
reaches the observed luminosity range, while the $35+5\,M_\odot$
product, used for the model atmospheres of
Section~\ref{subsec:ModAtm}, lies only $0.02$ to $0.10$\,dex below
the faintest observed star.

The surface composition provides an important additional distinction
between the models.
\new{In all four products, the companion material is deposited near the
hydrogen burning shell (Section~\ref{sec:merger_products}). The flash
mixes some of its hydrogen inward, where it burns. Winds remove most of
the remaining hydrogen rich material. In the two most massive products,
most of this wind stripping occurs during core helium burning.}
In the $35+5\,M_\odot$ product, winds progressively
remove the hydrogen rich layers and expose CNO processed material without
requiring rotationally induced mixing. The surface hydrogen abundance
enters the range $X_{\rm H}\simeq0.2$ to $0.55$ observed among the
apparently single SMC WN stars \citep{2015A&A...581A..21H} after about
$4.3\times10^4$\,yr, and the surface nitrogen abundance exceeds our
adopted WN threshold of $X_{\rm N}=2.25\times10^{-3}$ shortly afterwards,
at about $4.4\times10^4$\,yr. The luminosity and temperature criteria are
already satisfied by this time. The product therefore satisfies all three
adopted WN criteria for about $6.9\times10^5$\,yr, or $94\%$ of the
calculated post-merger evolution. The $40+2\,M_\odot$ product also reaches
the WN criteria, but its surface nitrogen abundance exceeds the adopted
threshold later, after about $2.6\times10^5$\,yr. Its luminosity and
temperature criteria are already satisfied by this
time. The product then satisfies all three WN criteria for about
$2.3\times10^5$\,yr. \new{This phase covers the final} $46\%$ of core helium
burning and \new{continues} until the end of the calculation,
\new{during core silicon burning}.

During most of the WN phase of the $35+5\,M_\odot$ product, its surface
hydrogen abundance lies between about $0.26$ and $0.38$, placing the
$X_{\rm H}=0.35$ adopted for the upper luminosity estimate toward the
hydrogen rich end of the calculated range. The surface hydrogen abundance
eventually declines to about $0.15$, while the total mass decreases from
about $18.8$ to $10.8\,M_\odot$ through wind mass loss.
The product remains relatively cool for much of the WN phase, with $T_{\rm eff}$
mostly between $23\,000$ and $33\,000$\,K, and reaches
$T_{\rm eff}>75\,000$\,K only during the final $\simeq5\times10^3$\,yr of
the calculation. In contrast, the $40+2\,M_\odot$ product is already at
$T_{\rm eff}\simeq95\,000$\,K when it first satisfies all three WN
criteria and remains hotter than this throughout its WN phase. Its surface
hydrogen abundance meanwhile falls from $0.25$ to $0.14$ and lies mostly
between $0.18$ and $0.23$, below the adopted $X_{\rm H}=0.35$ and partly
below the observed range. If rotational mixing were efficient in the
merger product, it could bring CNO processed material to the surface
earlier and make the surface nitrogen rise sooner than in these non
rotating models. It could also mix fresh hydrogen into the burning region,
so its net effect on the duration of the hydrogen rich WN phase is less
clear.

The two lower mass products show why luminosity alone is not sufficient to
identify a WN outcome. By core helium exhaustion, the maximum surface
nitrogen abundances of the $30+5$ and $20+2\,M_\odot$ products are only
$1.89\times10^{-3}$ and $1.90\times10^{-3}$, respectively, about $84\%$
of the adopted WN threshold and about $70\%$ of the abundance reached by
full CNO processing of our model mixture
(Section~\ref{sec:wr_class}). In contrast, the $35+5$ and
$40+2\,M_\odot$ products reach $X_{\rm N}=3.30\times10^{-3}$ and
$2.56\times10^{-3}$, respectively, at core helium exhaustion. The
$30+5\,M_\odot$ product exceeds the nitrogen threshold only after central
helium exhaustion, when renewed stripping also makes it hot. All three WN
criteria are then satisfied simultaneously for only about
$2.1\times10^3$\,yr, or $0.3\%$ of its calculated evolution. The
$20+2\,M_\odot$ product also reaches the nitrogen threshold only after
helium exhaustion, but its luminosity remains below
$\log L_{\rm min,WN}/L_\odot=5.33$, so it never satisfies the full WN
criteria. None of the four products approaches the adopted WC
composition.
The maximum surface $\mathrm{C/He}$ number ratio remains below
$10^{-3}$, compared with the adopted requirement
$\mathrm{C/He}\gtrsim0.1$.
This is expected: all four remnants are in core helium burning at
the merger, so the products of helium burning remain beneath the hydrogen
burning shell. The explosive helium shell route of
Section~\ref{sec:merger_products}, which requires a remnant with a carbon
core, is not represented among the products followed here.

\new{By core silicon burning, the three more massive products retain
only a thin layer of hydrogen bearing material. Although their surface
hydrogen mass fractions are still about $0.15$, each contains less
than $0.01\,M_\odot$ of hydrogen. Judged by their hydrogen content
alone, these stars are closer to Type~Ib than to Type~IIb supernova
progenitors. Their carbon oxygen cores range from $7$ to $12\,M_\odot$,
far more massive than those of the Type~IIb progenitors modelled by
\citet{2019MNRAS.490....9L}. We therefore do not assign them a
supernova type.

The closest comparison with those progenitors is our $20+2\,M_\odot$
product, which never reaches the WN phase in our calculation. At core
carbon burning it contains at most $0.04\,M_\odot$ of hydrogen. This
upper bound lies within the range that \citet{2019MNRAS.490....9L}
found at core collapse, although our calculation ends earlier.}

The full calculations also test the luminosity prescriptions used for the
grid. Considered together, the four models show a clear difference between
a relation evaluated using the core mass and one based on the total
remnant mass. The $40+2\,M_\odot$ product is brighter than the
$35+5\,M_\odot$ product by about $0.1$\,dex during core helium burning,
although its merged mass is slightly smaller, because its core is more
massive, $M_{\rm cp}=16.81$ compared with $13.95\,M_\odot$. Relative to the
relation of \citet{2011A&A...535A..56G}, evaluated at the contemporaneous
helium core mass as an approximate analogue of $M_{\rm cp}$, the models are
brighter by about $0.09$ to $0.52$\,dex during core helium burning. This
excess is supplied in large part by the hydrogen burning shell, which is
absent from a chemically homogeneous helium star and can dominate the
surface luminosity in partially stripped configurations
\citep{2025A&A...703A..40S}. The two prescriptions therefore differ much
more through the mass at which they are evaluated than through the fitted
relations themselves, which differ by less than $0.08$\,dex at equal mass
over our $M_{\rm cp}$ range. The comparison with the core mass relation is
also approximate because $M_{\rm cp}$ includes the hydrogen burning shell,
whereas the operational helium core mass does not. In the $35+5$ and
$40+2\,M_\odot$ models, the latter is also strongly affected by the
flash driven redistribution of hydrogen.

The relations of \citet{2025A&A...703A..40S}, evaluated at the
contemporaneous total mass, describe the four products more closely, but
do not bracket them at all epochs. While the $35+5$, $30+5$, and
$20+2\,M_\odot$ products still retain most of their hydrogen rich
envelopes, they are fainter than a pure helium star of the same total mass
because their helium cores are substantially smaller than $M_{\rm tot}$.
The $35+5$ and $30+5\,M_\odot$ products therefore lie below the pure
helium relation for most of core helium burning, typically by about
$0.1$\,dex and by at most $0.16$\,dex, and cross above it only as the
envelope is removed. In the $30+5\,M_\odot$ model, this crossing occurs at
$12.2\,M_\odot$, shortly before core helium exhaustion, after which the
model lies up to $0.39$\,dex above the relation; in the $35+5\,M_\odot$
model it occurs between $15.7$ and $12.9\,M_\odot$. The $20+2\,M_\odot$
product, which retains a smaller envelope fraction, follows the same
pattern much more weakly: it falls at most $0.03$\,dex below the relation
early in core helium burning and rises to $0.19$\,dex above it by helium
exhaustion. The $40+2\,M_\odot$ product behaves differently. Its core makes
up a larger fraction of its total mass, $M_{\rm cp}/M_{\rm tot}=0.90$
compared with $0.69$ to $0.76$ for the other three, and after the first
$10^4$\,yr it lies above the pure helium relation by about $0.02$ to
$0.16$\,dex for the rest of core helium burning. During core helium
burning, all four models remain below or within $0.01$\,dex of the maximum
luminosity relation evaluated for $X_{\rm H}=0.35$. The largest excursions
above this relation, about $0.11$ and $0.13$\,dex, occur during the post
exhaustion contraction of the $30+5$ and $40+2\,M_\odot$ models,
respectively.

The two total mass relations therefore give a useful luminosity range for
the grid, but not one that brackets a merger product at all epochs. The
models are never fainter than a helium star evaluated at their own helium
core mass, as the comparison above shows. The $35+5$, $30+5$, and
$20+2\,M_\odot$ products fall below the pure helium relation while their
total masses substantially exceed their core masses, but the
$40+2\,M_\odot$ product has a much larger core mass fraction and lies above
this relation for most of core helium burning. As the envelope is removed
and the core and total masses converge, the other three models also rise
above the pure helium relation. The two prescriptions of
Section~\ref{sec:wr_class} nevertheless bracket the calculated luminosity
during most of core helium burning, with the core mass relation providing
the lower estimate and the maximum total mass relation the upper estimate.
The grid level luminosities based on the maximum total mass relation are
correspondingly upper estimates. The calculations also show, as above,
that luminosity by itself does not determine whether a merger product
develops the temperature and surface composition of a WN star.

\subsection{Properties of the merger population}
\label{sec:population}

The population considered here consists of the Companion mergers. In these
systems, the envelope has been removed down to $M_{\rm cp}$, the remnant
fits within its Roche lobe at $a_{\rm f}$, and the stripped companion of
mass $M_2^*$ subsequently merges with it. Remnant contact mergers are
excluded, as discussed in Section~\ref{sec:outcomes}, because no detached
remnant of mass $M_{\rm cp}$ can be established at the inferred
$a_{\rm f}$ and our prescription therefore does not assign a unique merged
mass.
Under the enthalpy criterion, the $1055$ Companion mergers have remnant
masses $M_{\rm cp}=6.2$ to $19.6\,M_\odot$ and merged masses
$M_{\rm tot}=M_{\rm cp}+M_2^*=7.6$ to $24.6\,M_\odot$. Under the standard
criterion, the $187$ Companion mergers span $M_{\rm cp}=6.2$ to
$18.1\,M_\odot$ and $M_{\rm tot}=8.6$ to $22.9\,M_\odot$. The compression
point is a property of the pre-CE donor and does not depend on the
criterion. The different mass ranges arise primarily because the criteria
select different systems as Companion mergers, although the amount of
companion stripping also differs.
\new{During the inspiral, the companion loses less than $8\%$ of its mass,
with a median loss of about $1\%$ under the enthalpy criterion and $3\%$
under the standard criterion. This material joins the ejected envelope.
The remaining companion mass, $M_2^*=0.94$ to $5.0\,M_\odot$, merges with
the remnant.}
Under the standard criterion, none of the $45\,M_\odot$ donors and
only two of the $42.5\,M_\odot$ donors \new{undergo a Companion merger}.

The WN luminosity threshold of Equation~\ref{eq:Lmin} translates into a
characteristic mass for each luminosity prescription of
Section~\ref{sec:wr_class}. For the helium star relation evaluated at
$M_{\rm cp}$, it corresponds to $M_{\rm cp}=12.5\,M_\odot$. For the
relations evaluated at the total mass, it is reached at
$M_{\rm tot}=12.5\,M_\odot$ for the pure helium relation and at
$M_{\rm tot}=9.0\,M_\odot$ for the maximum luminosity relation at
$X_{\rm H}=0.35$. Merger products between these two masses therefore occupy
the regime in which whether they satisfy the WN luminosity criterion
depends most strongly on the contribution from the retained hydrogen rich
material and its burning shell.

Figure~\ref{fig:luminosityS25} compares the two total mass relations with
the four explicitly evolved merger products of
Section~\ref{sec:merger_evolution}, while
Figure~\ref{fig:parameter_space_outcomes} shows how the same luminosity
criteria map onto the full enthalpy grid. 
The pure helium relation
evaluated at $M_{\rm tot}$ gives $\log L/L_\odot=4.9$ to $5.8$ and places
about $74\%$ of the Companion mergers above $L_{\rm min,WN}$. These
systems first appear among $25\,M_\odot$ primaries, generally for
companions of $4\,M_\odot$ or more, although the exceptional
$1400\,R_\odot$ donor discussed above also exceeds the threshold for
$2$ and $2.5\,M_\odot$ companions. Among the $27.5\,M_\odot$ primaries,
all Companion mergers with companions of $3\,M_\odot$ or more exceed
the threshold, as do some with $2.5\,M_\odot$ companions and the two
systems with $1$ and $1.5\,M_\odot$ companions at $1600\,R_\odot$.
From $30\,M_\odot$ upwards every
Companion merger exceeds the threshold. The maximum luminosity relation
broadens this region substantially. It gives $\log L/L_\odot=5.2$ to $5.9$
and places about $96\%$ of the population above $L_{\rm min,WN}$. The
systems that remain below the threshold are the $20\,M_\odot$ donors with
companions of $1.5$ to $2.5\,M_\odot$ and the $22.5\,M_\odot$ donors with
$1.5\,M_\odot$ companions, together about $4\%$ of the population. Even
these exceed $L_{\rm min,WC}$.

The helium star relation evaluated at $M_{\rm cp}$ gives a more restrictive
picture because it uses the core mass rather than the total merged mass. It
spans $\log L/L_\odot=4.8$ to $5.7$ and places about $54\%$ of the
Companion mergers above $L_{\rm min,WN}$. The qualifying systems are
confined to primaries of $32.5\,M_\odot$ and above, where every Companion
merger exceeds the threshold.
The explicit calculations of
Section~\ref{sec:merger_evolution} place the actual merger products between
the core mass and maximum total mass estimates during most of core helium
burning. The core mass relation underestimates their luminosities, while
the maximum total mass relation remains an upper estimate during most of
this phase.

The same luminosity thresholds apply under the standard energy criterion,
but to a smaller population concentrated toward larger nominal Roche lobe
radii, where sufficient orbital energy is available to remove the envelope
down to $M_{\rm cp}$ while leaving the remnant within its Roche lobe
(Section~\ref{sec:results_outcomes} and
Figure~\ref{fig:parameter_space_outcomes_std}). The pure helium estimate
places about $60\%$ of these $187$ systems above $L_{\rm min,WN}$, the
maximum luminosity estimate $99\%$, and the core mass relation $20\%$.
With the pure helium relation, systems above $L_{\rm min,WN}$ first
appear among $25\,M_\odot$ primaries, and every Companion merger with
a primary of $27.5\,M_\odot$ or above exceeds the threshold.
With the maximum luminosity relation, the qualifying region begins
already at $20\,M_\odot$ for companions of $3\,M_\odot$ or more
and includes every Companion merger with a primary of $22.5\,M_\odot$
or above.

These luminosity based estimates should not be interpreted directly as
predicted fractions of Companion merger products that would be identified
as WN stars, for two reasons. First, the fraction obtained from the maximum
total mass relation is optimistic, since this relation provides an upper
luminosity estimate during most of core helium burning, as discussed above.
The pure helium relation does not provide such an upper estimate and can
instead underestimate the luminosity of a merger product with a large core
mass fraction. Second, luminosity is only a necessary condition for a WN
classification, while the effective temperature and surface composition
can presently be tested only for the explicitly evolved products.

None of the four explicitly evolved products approaches the adopted WC composition.
The explosive helium shell mechanism discussed in
Section~\ref{sec:merger_products} requires a remnant that has
already formed a carbon core.
None of our donors has reached this stage at Roche lobe overflow.
We do not construct a merger product from the most evolved donor,
the $25\,M_\odot$ star at $1400\,R_\odot$.
The $35+5\,M_\odot$ product satisfies the luminosity, temperature, and
nitrogen criteria for about $6.9\times10^5$\,yr, or $94\%$ of its
calculated post-merger evolution.
The $40+2\,M_\odot$ product satisfies
all three WN criteria for about $2.3\times10^5$\,yr, covering the remaining
$46\%$ of core helium burning and continuing until the end of the
calculation, \new{during core silicon burning}. The
$30+5\,M_\odot$ product remains below the adopted nitrogen threshold
throughout core helium burning and also becomes cooler than $20\,000$\,K
late in that phase. After helium exhaustion, stripping exposes more
strongly CNO processed material while the star contracts, and all three WN
criteria are satisfied simultaneously for only about
$2.1\times10^3$\,yr, or $0.3\%$ of its calculated evolution. The
$20+2\,M_\odot$ product never reaches the WN luminosity threshold
(Section~\ref{sec:merger_evolution}). Taken together with the two reasons
above, these examples show that the luminosity based fractions derived for
the full grid should not be interpreted as predicted WN fractions.

\subsection{Properties of the surviving binaries}
\label{sec:survivors}

The $51$ binaries that survive the common envelope in
Section~\ref{sec:results_outcomes} consist of $50$ Post-CE binaries and one
classified as Reverse TTMT. Apart from the exceptional donor, they
contain primaries of $30\,M_\odot$ and above with
companions of $4.5$ to $5\,M_\odot$. The exceptional donor extends the
companion masses down to $3\,M_\odot$. The systems emerge from the CE at
$a_{\rm f}=5.1$ to $9.8\,R_\odot$, corresponding to orbital periods of
about $0.28$ to $0.90$\,d. The companions retain more than $99\%$ of their
mass and are therefore essentially unstripped main sequence stars, while
the remnants have $M_{\rm cp}=10.6$ to $19.6\,M_\odot$. At the pure helium
luminosity estimate, the remnants have $\log L/L_\odot=5.2$ to $5.7$ and
$45$ of the $51$ exceed $L_{\rm min,WN}$. Five of the six systems
below this threshold come from the exceptional donor. Its remnant
is near the end of core helium burning and is therefore only
approximately described by these relations.
At the maximum
luminosity estimate all of them exceed the threshold. For these remnants
$M_{\rm tot}=M_{\rm cp}$, so the helium star relations of
\citet{2011A&A...535A..56G} and \citet{2025A&A...703A..40S} are evaluated
at the same mass and differ by less than $0.02$\,dex.

The companions are much fainter than the remnants. Their interior
luminosities at $m=M_2^*$ agree with the thermal equilibrium luminosities
of stars of the same mass to better than $0.2$\,dex, giving
$\log L_2/L_\odot=1.9$ to $2.8$. The resulting bolometric luminosity ratios
are only $L_2/L_1=4\times10^{-4}$ to $2.3\times10^{-3}$ when the
remnant luminosity is taken from the pre-CE profile at $M_{\rm cp}$.
With the pure helium relation, the ratio reaches at most
$4\times10^{-3}$. The predicted edge-on radial velocity
semi-amplitudes of the stripped remnants are $K_1=150$ to
$240\,{\rm km\,s^{-1}}$, so the systems would nevertheless be conspicuous
as spectroscopic binaries.

These semi-amplitudes are well above the sensitivity of existing RV
searches in the SMC. \citet{Schootemeijer_2024} obtained six VLT-UVES
epochs over $1.5$\,yr for each of the seven apparently single SMC WR stars
and found peak-to-peak variations of only $\Delta \mathrm{RV}=6$ to
$23\,{\rm km\,s^{-1}}$, with a median
$\sigma_{\mathrm{RV}}\approx5\,{\rm km\,s^{-1}}$. For six randomly
sampled epochs and a circular orbit, the expected peak-to-peak variation is
about $1.68\,K_1$, so these measurements correspond approximately to
projected semi-amplitude limits of $K_1\lesssim4$ to
$14\,{\rm km\,s^{-1}}$. The Monte Carlo analysis of
\citet{Schootemeijer_2024}, which accounts for the actual observing epochs
as well as random orbital inclinations and phases, gives probabilities
below about $5\%$ for an individual target to hide a companion more massive
than approximately $5\,M_\odot$ with a period below one year (Fig.\,\ref{fig:Schoot_Constraints}).
For our
systems, reducing the predicted edge-on semi-amplitudes of $150$ to
$240\,{\rm km\,s^{-1}}$ below the observational limits would require
inclinations of only about $i\lesssim1^\circ$ to $5^\circ$, corresponding
to roughly $0.01\%$ to $0.4\%$ of random orientations. Although our $3$ to
$5\,M_\odot$ companions extend below the nominal mass sensitivity quoted by
\citet{Schootemeijer_2024}, their periods of about a day or less and large
predicted semi-amplitudes make them particularly favourable for RV
detection (Fig.~\ref{fig:Schoot_Constraints}).

Detecting the companion spectrum itself would be much harder.
\citet{Schootemeijer_2024} note that even in known SMC WR binaries with
O-star companions of tens of solar masses, the He\,{\sc i} $\lambda4471$
absorption reaches only $0.02$ to $0.1$ of the continuum, and argue that
radial velocities therefore constrain faint, low mass companions far better
than searches for their spectral features. The much less massive companions
in our surviving systems contribute below $0.4\%$ of the bolometric
luminosity and would therefore be considerably more difficult to detect
directly. The expected observational signature is thus that of a
single-lined spectroscopic binary, with the companion revealed through the
reflex motion of the WR star rather than through its own spectrum. These
systems therefore do not provide a significant channel for producing WN
stars that would be classified as single by an RV survey with the
sensitivity of \citet{Schootemeijer_2024}.

\subsection{Comparison with model atmospheres}
\label{subsec:ModAtm}

The criteria implemented in this work to classify models into WR stars are based on prescriptions calibrated to entire populations and do not necessarily apply to individual models. We therefore compute detailed model atmospheres for a representative merger model at distinct temperatures and examine whether their spectral appearance resembles that of WR stars observed in the SMC. For this purpose, we choose the $35 + 5\,M_\odot$ merger model, since it lies roughly in the regime of the lowest luminosity SMC WR stars, which are apparently single.
We compute model atmospheres at two temperatures representative of the
cool and hot ends of the range observed among SMC WN stars: $\approx 45\,$kK (\object{SMC~AB~4}) and $\approx 100\,$kK (e.g., \object{SMC~AB~10}). The luminosity $\log L$, mass $M_*$, and elemental surface mass fractions exhibited by our $35 + 5\,M_\odot$ merger model in the vicinity of these two temperatures are given in Table\,\ref{tab:merger_models}.

For the computation of the models, we use the Potsdam Wolf-Rayet (PoWR) model atmosphere code \citep{Graefener2002, Hamann2003, Sander2015}, which solves the radiative transfer in the expanding atmospheres of hot stars without assuming local thermodynamic equilibrium (NLTE). The subsonic velocity field is obtained from the continuity equation assuming hydrostatic equilibrium, while the supersonic regime is stitched at the sonic point adopting a standard $\beta$-law with $\beta = 1$ for the velocity field \citep{Sander2015}. The inner radius of the model is defined at a mean Rosseland optical depth of $\tau_{\rm Ross} = 20$, where the surface effective temperature $T_*$, radius $R_*$, and stellar luminosity $L$ are defined and are related via the Stefan-Boltzmann equation.  For simplicity, we only consider model atoms of H, He, C, N, and Fe (including iron-group elements such as Sc, Ti, Ni). The iron abundance corresponds to the metallicity of $Z = 0.004$ adopted for the merger models. 

Beyond the surface elemental mass fractions $X_{\rm H}$, $X_{\rm He}$, $X_{\rm C}$, $X_{\rm N}$, $X_{\rm Fe}$, the input parameters needed to compute the models are $T_*$, $\log L$ (or $R_*$), the stellar mass $M_*$, and the wind parameters. Decisive for the WR appearance of the star is the mass loss rate $\dot M$.  In general, the winds of stripped products of WR stars are subject to large uncertainties \citep{Gotberg2018, Shenar_2020, Vink2017, 2025A&A...703A..40S}. Various prescriptions exist in the literature. For WR stars, both theoretical \citep{Sander2020, 2025A&A...703A..40S, Sabhahit2026} as well as empirical \citep{Nugis2000, 2019A&A...627A.151S}
exist in the literature. For hot stripped stars with optically thin winds, mainly theoretical prescriptions exist \citep[e.g.][]{Vink2017}, with very few observational counterparts \citep{Drout2023, Gotberg2023}. The most advanced theoretical prescriptions \citep[e.g.][]{Sabhahit2026} are not applicable at arbitrary temperatures and metallicities. We therefore opt for a pragmatic approach, and adopt two extremes, selecting a high $\dot{M}$ from empirical WR prescriptions by \citet{2019A&A...627A.151S, Shenar2020_Erratum} and a low $\dot{M}$ from theoretical $\dot{M}$ prescriptions for stripped stars with optically thin winds by \citet{Vink2017}. These span close to an order-of-magnitude, and hence probe the uncertainty in $\dot{M}$ to a sufficient degree in the context of our work. The adopted $\dot{M}$ values are given in Table\,\ref{tab:merger_models}.

For simplicity, we adopt terminal velocities for the cool and hot models corresponding to the measured values of SMC~AB~4 and SMC~AB~10, namely, $1000$ and $2000\,{\rm km\,s^{-1}}$, respectively. Clumping is accounted for via the microclumping approach, which assumes that the radiation forms in optically thin clumps that are a factor $D$ denser compared to the equivalent smooth wind. We adopt a fixed value of $D=10$, as inferred observationally \citep{2015A&A...581A..21H, 2016A&A...591A..22S}. The formal integration of the radiative transfer in the observer's frame is performed assuming a fixed microturbulent velocity of $\varv_{\rm mic} = 100\,{\rm km\,s^{-1}}$.

\begin{table}
\centering
\caption{Adopted parameters of the $35+5\,M_\odot$ merger models.}
\label{tab:merger_models}
\begin{tabular}{lcc}
\hline
Parameter & cool model & hot model \\
\hline
$T_*$ [kK] 
    & 45.1 
    & 100.1 \\

$\log (L/L_\odot)$ 
    & 5.51 
    & 5.48 \\

$M_*$ [$M_\odot$] 
    & 11.329 
    & 10.95 \\

$X_{\rm H}$ 
    & 0.26 
    & 0.25 \\

$X_{\rm N}$ 
    & $3.33\times10^{-3}$ 
    & $3.31\times10^{-3}$ \\

$X_{\rm C}$ 
    & $3.36\times10^{-5}$ 
    & $4.68\times10^{-5}$ \\

$X_{\rm Fe}$ 
    & $4.2\times10^{-4}$ 
    & $4.2\times10^{-4}$  \\

$\varv_\infty$ [km\,s$^{-1}$] 
    & 1000 
    & 2000 \\

$\log \dot{M}_{\rm low}$ [$M_\odot\,{\rm yr}^{-1}$]
    & -6.2 
    & -6.25 \\

$\log \dot{M}_{\rm high}$ [$M_\odot\,{\rm yr}^{-1}$]
    & -5.2 
    & -5.5 \\
\hline
\end{tabular}
\end{table}

\begin{figure}
    \centering
    \includegraphics[width=1.0\linewidth]{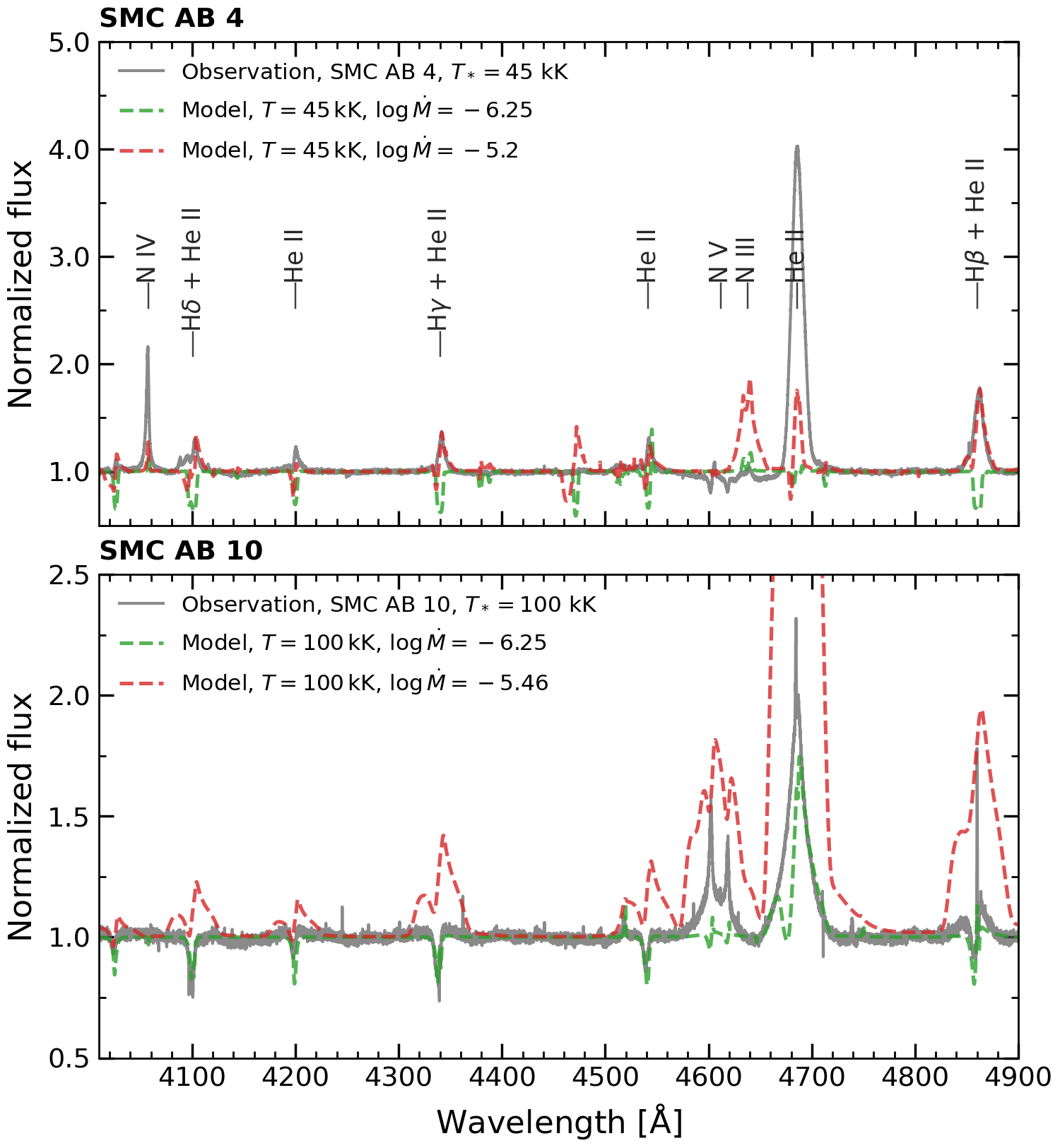}
\caption{
Comparison between co-added UVES observations \citep{Schootemeijer_2024} of the cool (SMC~AB~4, $T_* = 45\,$kK) and hot (SMC~AB~10, $T_* = 100\,$kK) SMC WR stars (gray solid line) with model atmospheres computed for the representative $35 + 5\,M_\odot$ merger models described in Table\,\ref{tab:merger_models}. Each model was computed with a low and high mass loss rate (see text for detail). Overall, all merger models presented here but the low $\dot{M}$ $45\,$kK model would be classified as WR. The models show a plausible agreement with observations, considering they have not been produced in order to fit these specific objects, and wind parameters have not been tuned.
}
    \label{fig:ModAt}
\end{figure}

In Fig.\,\ref{fig:ModAt}, we compare the synthetic spectra obtained for the cool and hot $35 + 5\,M_\odot$ merger models
with co-added optical spectra presented in \citet{Schootemeijer_2024} in the range $4\,000$ to $5\,000$\,\AA.
The spectra were acquired with the Ultraviolet and Visual Echelle Spectrograph (UVES) mounted on the Very Large Telescope (VLT) in Chile (ID: 108.22M1.001; PI: Schootemeijer). They have a resolving power of $R\approx 42\,000$ and a combined signal-to-noise (S/N) of $\approx 100$ per pixel. 

The low $\dot{M}$ models for both $T_*=$45\,kK and 100\,kK  are marginally consistent with a WR spectrum, showing some He\,{\sc ii}\,$4686$ emission and marginal H$\beta$ emission. Such an appearance is consistent with the lowest luminosity WR stars in the SMC, such as SMC~AB~4 and SMC~AB~10, and also reminiscent of the low luminosity WN3/O3 stars identified in the LMC \citep{Neugent2017}. Unsurprisingly, the high-$\dot{M}$ models exhibit unambiguous WR spectra. Considering that the models were not computed to fit the data, the general morphological agreement between these models and observations suggests that merger models may indeed provide a viable route for reproducing the properties of single WR stars in the SMC.

\section{Conclusions}
\label{sec:conclusions}

We have explored common envelope interactions between $20$ to $45\,M_\odot$
primaries and $1$ to $5\,M_\odot$ companions at $Z=0.004$, asking whether
they can account for apparently single WN stars at low metallicity. The two
outcomes that we follow beyond the common envelope phase have opposite
observational implications. The rare systems that survive as binaries would
be readily identified through the radial velocities of the WR star and
therefore cannot account for the apparently single population. Companion
mergers, by contrast, are far more common than surviving binaries in our
grid, and are the dominant outcome under the enthalpy criterion. They leave
no companion to detect, while our explicit calculations show that their
products can become hot, luminous, hydrogen poor stars with surface nitrogen
abundances toward the lower end of the range derived for the observed WN
stars. This channel is therefore a plausible origin for apparently single
WN stars.

The existence of this channel over a substantial region of parameter space
does not depend on the adopted envelope ejection criterion, although its
frequency does.
The energy condition for removing the envelope down to the
compression point is satisfied in about $92\%$ of the models under the
enthalpy criterion and about $53\%$ under the standard criterion. Companion
mergers account for about $74\%$ of the grid under the enthalpy criterion
and about $13\%$ under the standard one, while surviving binaries account
for only about $4\%$ under the former and are absent under the latter.
In most of the cases where the energy condition is satisfied but a Companion
merger does not occur, the remnant itself overfills its Roche lobe at the
final separation. The two criteria therefore differ substantially in
how frequently the channel occurs, but both allow the same outcome:
the low mass companion helps remove the envelope and then merges
with the exposed remnant, rather than with an intact star.

More generally, the outcomes form a continuous progression in envelope
removal and the final configuration of the components, and our grid
spans this progression rather than populating only its extremes.
At one end, a Complete merger retains a substantial part of the
envelope, although some material may already have been ejected.
With increasing envelope removal, the companion
may reach and merge with the donor core while part of the envelope above
$M_{\rm cp}$ remains bound. When the energy condition for removing the
envelope down to $M_{\rm cp}$ is satisfied, the outcome depends on whether
the remnant and the stripped companion can fit within their Roche lobes.
The result is a Remnant contact merger if the remnant cannot fit, a
Companion merger if the remnant fits but the stripped companion does not,
and a surviving binary if both fit.

The intermediate case of this sequence, in which the companion merges with
the core beneath a partly retained envelope, is not explicitly followed by
our prescription. The hydrogen rich material of the companion may be deposited
near the hydrogen burning shell, where the resulting hydrogen flash could
provide additional energy to eject the remaining envelope. Thus, even
systems for which our energy criterion predicts incomplete envelope
ejection may ultimately produce stripped merger products. Quantifying
this possibility requires a more detailed treatment of the merger and
subsequent hydrogen burning.

Surviving binaries remain confined to a small and favourable region of the
grid. Apart from the exceptional $25\,M_\odot$ donor at $1400\,R_\odot$,
which fills its Roche lobe near the end of core helium burning, they originate
from primaries of $30\,M_\odot$ and above with $4.5$ to $5\,M_\odot$
companions. The exceptional donor extends the companion masses down to
$3\,M_\odot$.
Surviving binaries occur only under the enthalpy criterion and account
for about $4\%$ of the models. They emerge at separations of $5.1$ to
$9.8\,R_\odot$, with orbital periods of about $0.28$ to $0.90$\,d and
predicted edge-on radial velocity semi-amplitudes of $150$ to
$240\,{\rm km\,s^{-1}}$ for the stripped remnants. Although the
companions contribute at most about $0.4\%$ of the bolometric luminosity
and would be difficult to detect directly, the reflex motion of the
remnant would be readily measurable. Using the sensitivity of the SMC RV
survey of \citet{Schootemeijer_2024}, hiding these systems would require
inclinations of only $i\lesssim1^\circ$ to $5^\circ$, corresponding to
roughly $0.01\%$ to $0.4\%$ of random orientations.

The merger products can readily be luminous enough to satisfy the WN
luminosity criterion. Under the enthalpy criterion, about $74\%$ exceed
$L_{\rm min,WN}$ using the pure helium relation evaluated at the total mass,
and about $96\%$ do so using the maximum luminosity relation for models
retaining hydrogen. The core mass estimate is more restrictive, placing
about $54\%$ above the threshold.
Qualifying products arise from primaries
of $25\,M_\odot$ and above at the pure helium estimate, and from
$20\,M_\odot$ at the maximum luminosity estimate. The explicit calculations
show that during most of core helium burning the merger luminosities lie
between the core mass and maximum total mass estimates. The core mass
relation underestimates their luminosities, while the maximum total mass
relation remains an upper estimate. The pure helium relation evaluated at
the total mass does not provide such an upper estimate and can lie either
below or above the calculated luminosity, depending on the core mass
fraction. The luminosity based fractions should therefore be interpreted as
fractions satisfying the adopted luminosity criterion rather than as
predicted WN fractions.

Luminosity alone does not determine the classification, as the four
explicitly evolved products show. The $35+5$ and $40+2\,M_\odot$ products
satisfy the luminosity, temperature, and surface nitrogen criteria
simultaneously for substantial parts of their core helium burning
evolution. The $30+5\,M_\odot$ product remains below the adopted nitrogen
threshold throughout core helium burning and satisfies all three WN
criteria only briefly after helium exhaustion. The $20+2\,M_\odot$ product
never reaches the WN luminosity minimum. None of the four products
approaches the adopted WC composition.

The adopted luminosity, temperature, and abundance criteria remain proxies
for a spectroscopic WN classification. In particular, the nitrogen threshold
is an observational reference rather than a sharp boundary. The emergent
spectrum also depends on the wind density and ionisation structure, which our
models do not predict. The PoWR model atmospheres of Section~\ref{subsec:ModAtm}, computed for the $35+5\,M_\odot$ product at $T_*=45$ and $100$\,kK, show unambiguous WR
spectra at the empirical WR mass loss rate and only marginal emission at the
theoretical rate for optically thin winds. Whether the product is observed as
a WR star is therefore set by its wind, which the evolutionary calculation
does not determine. 

The temperature comparison is more model dependent. The
$35+5\,M_\odot$ product spends most of its WN phase at
$T_{\rm eff}\simeq23\,000$ to $33\,000$\,K and exceeds
$75\,000$\,K only during the final $\simeq5\times10^3$\,yr of the
calculation. In contrast, the $40+2\,M_\odot$ product is already at
$T_{\rm eff}\simeq95\,000$\,K when it first satisfies all three WN
criteria and remains hotter thereafter, similar to the majority of the
observed SMC WN stars, which have $T_*>75\,000$\,K
\citep{2015A&A...581A..21H}. The observed $T_*$, and the prescribed $T_*$ of the model atmospheres
of Section~\ref{subsec:ModAtm}, is defined at a Rosseland
optical depth of 20 in the atmosphere models and is therefore not directly
comparable to the model $T_{\rm eff}$.

The fractions quoted above refer to our unweighted model grid and are not
predicted occurrence rates for the SMC. Their quantitative values are also
subject to the uncertainties inherent in common envelope modelling. The
donor structure at the onset of Roche lobe overflow sets the envelope binding
energy and compression point, and depends on uncertain mixing and wind mass
loss. The final separation follows from an energy balance that assumes all
released orbital energy is available for envelope ejection, while the
companion is treated as a point mass during the binary evolution and its
stripping during the inspiral is estimated from a ram pressure balance with
a fixed density reduction factor $f=10$, neglecting corotation. The surviving
systems retain more than $99\%$ of the companion mass, but their radii are
reduced to $0.68$ to $0.92$ of the unperturbed values. Without this reduction,
only four of the 51 systems would fit within their Roche lobes at $a_{\rm f}$.
The boundary between survival and Companion merger is therefore sensitive
to the adopted stripping prescription. The relative frequencies of the
merger outcomes remain sensitive to the adopted envelope ejection criterion.

A simple estimate of the relative numbers is instructive. The SMC
contains five known WR binaries and seven apparently single WR stars
\citep{Schootemeijer_2024}. The five binaries are also the most luminous
WR stars, while the five least luminous stars are apparently single
\citep{Schootemeijer_2018}. The latter have generally lower inferred
masses, supporting their formation from less massive progenitors.
The known WR binaries have progenitors generally of about
$60\,M_\odot$ or more \citep{2016A&A...591A..22S}, whereas our
Companion merger channel produces WN candidates from substantially
less massive primaries.

We compare merger progenitors of $30$ to $45\,M_\odot$ with binary
progenitors of $60$ to $100\,M_\odot$ and $M_2/M_1>0.5$, assuming a
uniform initial mass ratio distribution, a Salpeter initial mass
function, and initial separations distributed uniformly in $\log a$
over the sampled range. The initial progenitor populations have a
number ratio of about $0.48$. If the two populations have the same
efficiency of producing observable WR stars through common envelope
evolution, the five observed WR binaries would correspond to about
2.4 WN merger products. Extending the merger progenitors down to
$25\,M_\odot$ increases this estimate to about 4.5. This scaling
applies under either envelope ejection criterion, provided the
corresponding formation efficiencies are equal for both populations.
The actual ratio depends on their relative formation efficiencies,
which need not be equal even under the same energy criterion.

If the massive WR binaries instead form through stable mass transfer,
the comparison changes. Their formation efficiency is then determined
by the mass transfer stability criterion rather than the common
envelope energy prescription. The reduction in Companion mergers
under the standard energy criterion would therefore not be accompanied
by a corresponding reduction in WR binary formation.

The Companion mergers counted in our grid, however, may represent
only part of this channel. In a Remnant contact merger, the energy
budget is also sufficient to remove the envelope down to $M_{\rm cp}$,
but the remnant overfills its Roche lobe at the final separation.
Although we do not follow the subsequent coalescence, it may also
produce a similarly stripped merger product.
If all Remnant contact mergers contribute to the same channel, the
fraction of the grid forming such products increases from about
$13\%$ to $53\%$ under the standard criterion, a factor of four,
and from about $74\%$ to $88\%$ under the enthalpy criterion.
For similarly constructed merger products, the subsequent evolution
in one dimension is not expected to depend strongly on which of
these two merger paths produced them. The main uncertainty lies
in the three-dimensional coalescence, which determines how much
of the envelope is ultimately ejected and how the companion
material is incorporated into the remnant.

The relative numbers would also depend on the fraction of massive
binaries that undergo stable mass transfer and subsequently become
observable WR binaries. The estimates neglect differences in
observable lifetimes and binaries outside the sampled separation
range. A population synthesis that uses our outcomes and reproduces
the observed number of SMC WR binaries is needed to determine the
actual contribution of this channel.

The post-merger calculations carry a different set of uncertainties. We
construct the merger products by shock heating the disrupted companion
material, sorting it into the remnant by entropy, and relaxing the resulting
structure in one dimension without rotation. This construction is guided by
the two-dimensional simulations of the stream-core interaction of
\citet{2002MNRAS.334..819I} and the post-merger calculations of
\citet{Ivanova2003}, but the depth to which the companion material
penetrates, the heating and mixing it experiences, and the angular momentum
retained by the product are prescribed rather than calculated.
The time at which CNO processed material reaches the surface, and hence
the duration of the WN phase, also depends on the adopted wind mass loss,
here the \citet{Vink2001} rate and the super Eddington wind of
Section~\ref{sec:merger_products}, neither of which is established for
hydrogen poor stars at this metallicity.
The surface nitrogen enrichment of the $35+5\,M_\odot$ product in
particular follows from the flash driven mixing of hydrogen into the
helium burning core, a consequence of the instantaneous placement of the
companion material.
Determining the penetration, heating, and mixing of the companion
material requires three-dimensional simulations of a low mass star
merging with a stripped remnant undergoing core helium burning.
Such simulations would directly test our construction of the merger
products. The present calculations identify the relevant parameter space
and provide the stellar structures of the stripped remnants that can
serve as their initial conditions.

Low mass companions can strongly affect the evolution of massive stars at
low metallicity. When they survive the common envelope interaction, they
leave an unmistakable radial velocity signature. In the Companion merger
channel, which is much more common than binary survival, they instead help
remove the envelope and then merge with the exposed helium burning remnant.
The resulting star can become hot, luminous, hydrogen poor, and nitrogen
enriched. A low mass companion and a common envelope are therefore
essential to the stripping in this channel, yet the event is seen only as a
luminous red nova, and the star that remains has
no companion to detect and did not need to shed its envelope by winds
alone.

\begin{acknowledgments}
N.I. acknowledges funding from NSERC under Discovery grant No. RGPIN-2025-05603.
N.I. and T.S. thank the Aspen Center for Physics, where this work was initiated in part and which is supported by National Science Foundation grant PHY-2210452.
N.I. thanks Stephen Justham for fruitful discussions during a visit as a guest of the Excellence Cluster ORIGINS (EXC-2094, Project ID 390783311).
N.I. acknowledges the use of OpenAI's ChatGPT as an aid for language editing and of Anthropic's Claude for the Python script that draws objects in Figure~\ref{fig_scenario}.
This research was enabled in part by support provided by the University of Alberta Digital Research Services (ualberta.ca) and the Digital Research Alliance of Canada (alliancecan.ca). Computations were performed on the Trillium supercomputer cluster hosted by SciNet. N.I. and H.M. also thank Dmitry Pogosian for allowing them to use the Zeldovich computer cluster for part of the calculations. T.S. acknowledges support from the European Research Council (ERC) under the European Union's Horizon 2020 research and innovation programme (grant agreement 101164755/METAL).
\end{acknowledgments}

\bibliographystyle{aasjournalv7}
\bibliography{wr_ref}

\begin{thebibliography}{}
\expandafter\ifx\csname natexlab\endcsname\relax\def\natexlab#1{#1}\fi
\providecommand{\url}[1]{\href{#1}{#1}}
\providecommand{\dodoi}[1]{doi:~\href{http://doi.org/#1}{\nolinkurl{#1}}}
\providecommand{\doeprint}[1]{\href{http://ascl.net/#1}{\nolinkurl{http://ascl.net/#1}}}
\providecommand{\doarXiv}[1]{\href{https://arxiv.org/abs/#1}{\nolinkurl{https://arxiv.org/abs/#1}}}

% type= article
\bibitem[{D.~R. Aguilera-Dena {et~al.}(2022)Aguilera-Dena, Langer, Antoniadis,
  Pauli, Dessart, Vigna-Gómez, Gräfener, \& Yoon}]{Aguilera_Dena_2022}
Aguilera-Dena, D.~R., Langer, N., Antoniadis, J., {et~al.} 2022,
  \bibinfo{title}{Stripped-envelope stars in different metallicity
  environments: I. Evolutionary phases, classification, and populations,} A\&A,
  661, A60, \dodoi{10.1051/0004-6361/202142895}

% type= article
\bibitem[{C. {Angulo} {et~al.}(1999){Angulo}, {Arnould}, {Rayet},
  {Descouvemont}, {Baye}, {Leclercq-Willain}, {Coc}, {Barhoumi}, {Aguer},
  {Rolfs}, {Kunz}, {Hammer}, {Mayer}, {Paradellis}, {Kossionides}, {Chronidou},
  {Spyrou}, {degl'Innocenti}, {Fiorentini}, {Ricci}, {Zavatarelli},
  {Providencia}, {Wolters}, {Soares}, {Grama}, {Rahighi}, {Shotter}, \& {Lamehi
  Rachti}}]{Angulo1999}
{Angulo}, C., {Arnould}, M., {Rayet}, M., {et~al.} 1999, \bibinfo{title}{{A
  compilation of charged-particle induced thermonuclear reaction rates},}
  \nphysa, 656, 3, \dodoi{10.1016/S0375-9474(99)00030-5}

% type= article
\bibitem[{A. Ballone {et~al.}(2023)Ballone, Costa, Mapelli, MacLeod,
  Torniamenti, \& Pacheco-Arias}]{Ballone2023}
Ballone, A., Costa, G., Mapelli, M., {et~al.} 2023, \bibinfo{title}{Formation
  of black holes in the pair-instability mass gap: hydrodynamical simulations
  of a head-on massive star collision,} Monthly Notices of the Royal
  Astronomical Society, 519, 5191, \dodoi{10.1093/mnras/stac3752}

% type= article
\bibitem[{G.~S. {Bisnovatyi-Kogan} \& Y.~B. {Zeldovich}(1967){Bisnovatyi-Kogan}
  \& {Zeldovich}}]{BKZ67}
{Bisnovatyi-Kogan}, G.~S., \& {Zeldovich}, Y.~B. 1967,
  \bibinfo{title}{{Adiabatic Outflow from a Star and Equilibrium States with
  Excess Energy},} Sov. Astr., 10, 959

% type= article
\bibitem[{N. {Blagorodnova} {et~al.}(2017){Blagorodnova}, {Kotak}, {Polshaw},
  {et~al.}}]{Blagorodnova2017}
{Blagorodnova}, N., {Kotak}, R., {Polshaw}, J., {et~al.} 2017,
  \bibinfo{title}{{Common Envelope Ejection for a Luminous Red Nova in M101},}
  \apj, 834, 107, \dodoi{10.3847/1538-4357/834/2/107}

% type= article
\bibitem[{T. {Bloecker}(1995){Bloecker}}]{Blocker1995}
{Bloecker}, T. 1995, \bibinfo{title}{{Stellar evolution of low and
  intermediate-mass stars. I. Mass loss on the AGB and its consequences for
  stellar evolution.},} \aap, 297, 727

% type= article
\bibitem[{S. {Blouin} {et~al.}(2020){Blouin}, {Shaffer}, {Saumon}, \&
  {Starrett}}]{Blouin2020}
{Blouin}, S., {Shaffer}, N.~R., {Saumon}, D., \& {Starrett}, C.~E. 2020,
  \bibinfo{title}{{New Conductive Opacities for White Dwarf Envelopes},} \apj,
  899, 46, \dodoi{10.3847/1538-4357/ab9e75}

% type= article
\bibitem[{L. {Boco} {et~al.}(2025){Boco}, {Mapelli}, {Sander}, {Mesini},
  {Ramachandran}, {Torniamenti}, {Korb}, {Liu}, {Sabhahit}, \&
  {Vink}}]{2025A&A...703A.243B}
{Boco}, L., {Mapelli}, M., {Sander}, A. A.~C., {et~al.} 2025,
  \bibinfo{title}{{Metal-poor single Wolf-Rayet stars: The interplay of
  optically thick winds and rotation},} \aap, 703, A243,
  \dodoi{10.1051/0004-6361/202556187}

% type= article
\bibitem[{H. {Bondi} \& F. {Hoyle}(1944){Bondi} \&
  {Hoyle}}]{1944MNRAS.104..273B}
{Bondi}, H., \& {Hoyle}, F. 1944, \bibinfo{title}{{On the mechanism of
  accretion by stars},} \mnras, 104, 273, \dodoi{10.1093/mnras/104.5.273}

% type= article
\bibitem[{C. {Burkhardt} {et~al.}(2026){Burkhardt}, {Han}, {Oey}, {Ivanova}, \&
  {Renzo}}]{2026ApJ..1002..171B}
{Burkhardt}, C., {Han}, F., {Oey}, M.~S., {Ivanova}, N., \& {Renzo}, M. 2026,
  \bibinfo{title}{{Kinematics of Wolf─Rayet Stars in the LMC: Clues to
  Subtype Origins},} \apj, 1002, 171, \dodoi{10.3847/1538-4357/ae5492}

% type= article
\bibitem[{S. {Cassisi} {et~al.}(2007){Cassisi}, {Potekhin}, {Pietrinferni},
  {Catelan}, \& {Salaris}}]{Cassisi2007}
{Cassisi}, S., {Potekhin}, A.~Y., {Pietrinferni}, A., {Catelan}, M., \&
  {Salaris}, M. 2007, \bibinfo{title}{{Updated Electron-Conduction Opacities:
  The Impact on Low-Mass Stellar Models},} \apj, 661, 1094,
  \dodoi{10.1086/516819}

% type= article
\bibitem[{R.~A. {Chevalier} \& N. {Soker}(1989){Chevalier} \&
  {Soker}}]{1989ApJ...341..867C}
{Chevalier}, R.~A., \& {Soker}, N. 1989, \bibinfo{title}{{Asymmetric Envelope
  Expansion of Supernova 1987A},} \apj, 341, 867, \dodoi{10.1086/167545}

% type= article
\bibitem[{A.~I. {Chugunov} {et~al.}(2007){Chugunov}, {Dewitt}, \&
  {Yakovlev}}]{Chugunov2007}
{Chugunov}, A.~I., {Dewitt}, H.~E., \& {Yakovlev}, D.~G. 2007,
  \bibinfo{title}{{Coulomb tunneling for fusion reactions in dense matter: Path
  integral Monte Carlo versus mean field},} \prd, 76, 025028,
  \dodoi{10.1103/PhysRevD.76.025028}

% type= article
\bibitem[{P.~S. {Conti}(1975){Conti}}]{1975MSRSL...9..193C}
{Conti}, P.~S. 1975, \bibinfo{title}{{On the relationship between Of and WR
  stars.},} Memoires of the Societe Royale des Sciences de Liege, 9, 193

% type= article
\bibitem[{P.~A. {Crowther}(2007){Crowther}}]{2007ARAA..45..177C}
{Crowther}, P.~A. 2007, \bibinfo{title}{{Physical Properties of Wolf-Rayet
  Stars},} \araa, 45, 177, \dodoi{10.1146/annurev.astro.45.051806.110615}

% type= article
\bibitem[{P.~A. {Crowther} {et~al.}(2002){Crowther}, {Dessart}, {Hillier},
  {Abbott}, \& {Fullerton}}]{2002A&A...392..653C}
{Crowther}, P.~A., {Dessart}, L., {Hillier}, D.~J., {Abbott}, J.~B., \&
  {Fullerton}, A.~W. 2002, \bibinfo{title}{{Stellar and wind properties of LMC
  WC4 stars. A metallicity dependence for Wolf-Rayet mass-loss rates},} \aap,
  392, 653, \dodoi{10.1051/0004-6361:20020941}

% type= article
\bibitem[{P.~A. Crowther {et~al.}(2010)Crowther, Schnurr, Hirschi, Yusof,
  Parker, Goodwin, \& Kassim}]{Crowther2010}
Crowther, P.~A., Schnurr, O., Hirschi, R., {et~al.} 2010, \bibinfo{title}{The
  {R136} star cluster hosts several stars whose individual masses greatly
  exceed the accepted {150Msolar} stellar mass limit,} \mnras, 408, 731,
  \dodoi{10.1111/j.1365-2966.2010.17167.x}

% type= article
\bibitem[{R.~H. {Cyburt} {et~al.}(2010){Cyburt}, {Amthor}, {Ferguson},
  {Meisel}, {Smith}, {Warren}, {Heger}, {Hoffman}, {Rauscher}, {Sakharuk},
  {Schatz}, {Thielemann}, \& {Wiescher}}]{Cyburt2010}
{Cyburt}, R.~H., {Amthor}, A.~M., {Ferguson}, R., {et~al.} 2010,
  \bibinfo{title}{{The JINA REACLIB Database: Its Recent Updates and Impact on
  Type-I X-ray Bursts},} \apjs, 189, 240, \dodoi{10.1088/0067-0049/189/1/240}

% type= article
\bibitem[{M. {de Kool}(1990){de Kool}}]{1990ApJ...358..189D}
{de Kool}, M. 1990, \bibinfo{title}{{Common Envelope Evolution and Double Cores
  of Planetary Nebulae},} \apj, 358, 189, \dodoi{10.1086/168974}

% type= article
\bibitem[{A. {de Koter} {et~al.}(1997){de Koter}, {Heap}, \&
  {Hubeny}}]{1997ApJ...477..792D}
{de Koter}, A., {Heap}, S.~R., \& {Hubeny}, I. 1997, \bibinfo{title}{{On the
  Evolutionary Phase and Mass Loss of the Wolf-Rayet--like Stars in R136a},}
  \apj, 477, 792, \dodoi{10.1086/303736}

% type= article
\bibitem[{O. {De Marco} \& N. {Soker}(2002){De Marco} \&
  {Soker}}]{2002PASP..114..602D}
{De Marco}, O., \& {Soker}, N. 2002, \bibinfo{title}{{A New Look at the
  Evolution of Wolf-Rayet Central Stars of Planetary Nebulae},} \pasp, 114,
  602, \dodoi{10.1086/341691}

% type= article
\bibitem[{M.~R. {Drout} {et~al.}(2023){Drout}, {G{\"o}tberg}, {Ludwig}, {Groh},
  {de Mink}, {O'Grady}, \& {Smith}}]{Drout2023}
{Drout}, M.~R., {G{\"o}tberg}, Y., {Ludwig}, B.~A., {et~al.} 2023,
  \bibinfo{title}{{An observed population of intermediate-mass helium stars
  that have been stripped in binaries},} Science, 382, 1287,
  \dodoi{10.1126/science.ade4970}

% type= article
\bibitem[{R. {Edgar}(2004){Edgar}}]{2004NewAR..48..843E}
{Edgar}, R. 2004, \bibinfo{title}{{A review of Bondi-Hoyle-Lyttleton
  accretion},} \nar, 48, 843, \dodoi{10.1016/j.newar.2004.06.001}

% type= article
\bibitem[{P.~P. Eggleton(1983)Eggleton}]{eggleton_aproximations_1983}
Eggleton, P.~P. 1983, \bibinfo{title}{Aproximations to the radii of {Roche}
  lobes.,} \apj, 268, 368, \dodoi{10.1086/160960}

% type= article
\bibitem[{E. {Farrell} {et~al.}(2022){Farrell}, {Groh}, {Meynet}, \&
  {Eldridge}}]{Farrell2022}
{Farrell}, E., {Groh}, J.~H., {Meynet}, G., \& {Eldridge}, J.~J. 2022,
  \bibinfo{title}{{Numerical experiments to help understand cause and effect in
  massive star evolution},} \mnras, 512, 4116, \dodoi{10.1093/mnras/stac538}

% type= article
\bibitem[{J.~W. {Ferguson} {et~al.}(2005){Ferguson}, {Alexander}, {Allard},
  {Barman}, {Bodnarik}, {Hauschildt}, {Heffner-Wong}, \&
  {Tamanai}}]{Ferguson2005}
{Ferguson}, J.~W., {Alexander}, D.~R., {Allard}, F., {et~al.} 2005,
  \bibinfo{title}{{Low-Temperature Opacities},} \apj, 623, 585,
  \dodoi{10.1086/428642}

% type= article
\bibitem[{C. {Foellmi} {et~al.}(2003{\natexlab{a}}){Foellmi}, {Moffat}, \&
  {Guerrero}}]{2003MNRAS.338..360F}
{Foellmi}, C., {Moffat}, A.~F.~J., \& {Guerrero}, M.~A. 2003{\natexlab{a}},
  \bibinfo{title}{{Wolf-Rayet binaries in the Magellanic Clouds and
  implications for massive-star evolution - I. Small Magellanic Cloud},}
  \mnras, 338, 360, \dodoi{10.1046/j.1365-8711.2003.06052.x}

% type= article
\bibitem[{C. {Foellmi} {et~al.}(2003{\natexlab{b}}){Foellmi}, {Moffat}, \&
  {Guerrero}}]{2003MNRAS.338.1025F}
{Foellmi}, C., {Moffat}, A.~F.~J., \& {Guerrero}, M.~A. 2003{\natexlab{b}},
  \bibinfo{title}{{Wolf-Rayet binaries in the Magellanic Clouds and
  implications for massive-star evolution - II. Large Magellanic Cloud},}
  \mnras, 338, 1025, \dodoi{10.1046/j.1365-8711.2003.06161.x}

% type= article
\bibitem[{G.~M. {Fuller} {et~al.}(1985){Fuller}, {Fowler}, \&
  {Newman}}]{Fuller1985}
{Fuller}, G.~M., {Fowler}, W.~A., \& {Newman}, M.~J. 1985,
  \bibinfo{title}{{Stellar weak interaction rates for intermediate-mass nuclei.
  IV - Interpolation procedures for rapidly varying lepton capture rates using
  effective log (ft)-values},} \apj, 293, 1, \dodoi{10.1086/163208}

% type= article
\bibitem[{E. {Gaburov} {et~al.}(2008){Gaburov}, {Lombardi}, \& {Portegies
  Zwart}}]{2008MNRAS.383L...5G}
{Gaburov}, E., {Lombardi}, J.~C., \& {Portegies Zwart}, S. 2008,
  \bibinfo{title}{{Mixing in massive stellar mergers},} \mnras, 383, L5,
  \dodoi{10.1111/j.1745-3933.2007.00399.x}

% type= article
\bibitem[{A. Gilkis {et~al.}(2025)Gilkis, Laplace, Arcavi, Shenar, \&
  Schneider}]{10.1093/mnras/staf884}
Gilkis, A., Laplace, E., Arcavi, I., Shenar, T., \& Schneider, F. R.~N. 2025,
  \bibinfo{title}{The landscape of binary core-collapse supernova progenitors
  and the late emergence of Wolf–Rayet winds,} Monthly Notices of the Royal
  Astronomical Society, 540, 3094, \dodoi{10.1093/mnras/staf884}

% type= article
\bibitem[{Y. {G{\"o}tberg} {et~al.}(2018){G{\"o}tberg}, {de Mink}, {Groh},
  {Kupfer}, {Crowther}, {Zapartas}, \& {Renzo}}]{Gotberg2018}
{G{\"o}tberg}, Y., {de Mink}, S.~E., {Groh}, J.~H., {et~al.} 2018,
  \bibinfo{title}{{Spectral models for binary products: Unifying subdwarfs and
  Wolf-Rayet stars as a sequence of stripped-envelope stars},} \aap, 615, A78,
  \dodoi{10.1051/0004-6361/201732274}

% type= article
\bibitem[{Y. {G{\"o}tberg} {et~al.}(2023){G{\"o}tberg}, {Drout}, {Ji}, {Groh},
  {Ludwig}, {Crowther}, {Smith}, {de Koter}, \& {de Mink}}]{Gotberg2023}
{G{\"o}tberg}, Y., {Drout}, M.~R., {Ji}, A.~P., {et~al.} 2023,
  \bibinfo{title}{{Stellar Properties of Observed Stars Stripped in Binaries in
  the Magellanic Clouds},} \apj, 959, 125, \dodoi{10.3847/1538-4357/ace5a3}

% type= article
\bibitem[{G. {Gr{\"a}fener} {et~al.}(2002){Gr{\"a}fener}, {Koesterke}, \&
  {Hamann}}]{Graefener2002}
{Gr{\"a}fener}, G., {Koesterke}, L., \& {Hamann}, W.-R. 2002,
  \bibinfo{title}{{Line-blanketed model atmospheres for WR stars},} \aap, 387,
  244, \dodoi{10.1051/0004-6361:20020269}

% type= article
\bibitem[{G. {Gr{\"a}fener} {et~al.}(2011){Gr{\"a}fener}, {Vink}, {de Koter},
  \& {Langer}}]{2011A&A...535A..56G}
{Gr{\"a}fener}, G., {Vink}, J.~S., {de Koter}, A., \& {Langer}, N. 2011,
  \bibinfo{title}{{The Eddington factor as the key to understand the winds of
  the most massive stars. Evidence for a {\ensuremath{\Gamma}}-dependence of
  Wolf-Rayet type mass loss},} \aap, 535, A56,
  \dodoi{10.1051/0004-6361/201116701}

% type= article
\bibitem[{N. {Grevesse} \& A.~J. {Sauval}(1998){Grevesse} \& {Sauval}}]{GS1998}
{Grevesse}, N., \& {Sauval}, A.~J. 1998, \bibinfo{title}{{Standard Solar
  Composition},} \ssr, 85, 161, \dodoi{10.1023/A:1005161325181}

% type= article
\bibitem[{A. {Grichener} {et~al.}(2026){Grichener}, {Valli}, {Renzo}, {Mandel},
  {Kratter}, {Siwek}, {Sen}, \& {Shah}}]{2026arXiv261107326G}
{Grichener}, A., {Valli}, R., {Renzo}, M., {et~al.} 2026,
  \bibinfo{title}{{Double Neutron Star Mergers and Fast Blue Optical
  Transients: A Circumbinary Disk Connection},} arXiv e-prints,
  arXiv:2611.07326, \dodoi{10.48550/arXiv.2611.07326}

% type= article
\bibitem[{R. {Hainich} {et~al.}(2015){Hainich}, {Pasemann}, {Todt}, {Shenar},
  {Sander}, \& {Hamann}}]{2015A&A...581A..21H}
{Hainich}, R., {Pasemann}, D., {Todt}, H., {et~al.} 2015,
  \bibinfo{title}{{Wolf-Rayet stars in the Small Magellanic Cloud. I. Analysis
  of the single WN stars},} \aap, 581, A21, \dodoi{10.1051/0004-6361/201526241}

% type= article
\bibitem[{W.-R. {Hamann} \& G. {Gr{\"a}fener}(2003){Hamann} \&
  {Gr{\"a}fener}}]{Hamann2003}
{Hamann}, W.-R., \& {Gr{\"a}fener}, G. 2003, \bibinfo{title}{{A temperature
  correction method for expanding atmospheres},} \aap, 410, 993,
  \dodoi{10.1051/0004-6361:20031308}

% type= article
\bibitem[{W. {Hillebrandt} \& F. {Meyer}(1989){Hillebrandt} \&
  {Meyer}}]{1989A&A...219L...3H}
{Hillebrandt}, W., \& {Meyer}, F. 1989, \bibinfo{title}{{A common envelope
  model for SN 1987A.},} \aap, 219, L3

% type= article
\bibitem[{F. {Hoyle} \& R.~A. {Lyttleton}(1939){Hoyle} \&
  {Lyttleton}}]{1939PCPS...35..405H}
{Hoyle}, F., \& {Lyttleton}, R.~A. 1939, \bibinfo{title}{{The effect of
  interstellar matter on climatic variation},} Proceedings of the Cambridge
  Philosophical Society, 35, 405, \dodoi{10.1017/S0305004100021150}

% type= article
\bibitem[{J.~R. {Hurley} {et~al.}(2000){Hurley}, {Pols}, \&
  {Tout}}]{hurley_comprehensive_2000}
{Hurley}, J.~R., {Pols}, O.~R., \& {Tout}, C.~A. 2000,
  \bibinfo{title}{{Comprehensive analytic formulae for stellar evolution as a
  function of mass and metallicity},} \mnras, 315, 543,
  \dodoi{10.1046/j.1365-8711.2000.03426.x}

% type= article
\bibitem[{I. Iben \& M. Livio(1993)Iben \& Livio}]{iben_common_1993}
Iben, I., \& Livio, M. 1993, \bibinfo{title}{{COMMON} {ENVELOPES} {IN} {BINARY}
  {STAR} {EVOLUTION},} \pasp, 105, 1373, \dodoi{10.1086/133321}

% type= article
\bibitem[{C.~A. {Iglesias} \& F.~J. {Rogers}(1993){Iglesias} \&
  {Rogers}}]{Iglesias1993}
{Iglesias}, C.~A., \& {Rogers}, F.~J. 1993, \bibinfo{title}{{Radiative
  Opacities for Carbon- and Oxygen-rich Mixtures},} \apj, 412, 752,
  \dodoi{10.1086/172958}

% type= article
\bibitem[{C.~A. {Iglesias} \& F.~J. {Rogers}(1996){Iglesias} \&
  {Rogers}}]{Iglesias1996}
{Iglesias}, C.~A., \& {Rogers}, F.~J. 1996, \bibinfo{title}{{Updated Opal
  Opacities},} \apj, 464, 943, \dodoi{10.1086/177381}

% type= misc
\bibitem[{A.~W. {Irwin}(2004){Irwin}}]{Irwin2004}
{Irwin}, A.~W. 2004, The FreeEOS Code for Calculating the Equation of State for
  Stellar Interiors, \url{http://freeeos.sourceforge.net/}

% type= article
\bibitem[{N. {Itoh} {et~al.}(1996){Itoh}, {Hayashi}, {Nishikawa}, \&
  {Kohyama}}]{Itoh1996}
{Itoh}, N., {Hayashi}, H., {Nishikawa}, A., \& {Kohyama}, Y. 1996,
  \bibinfo{title}{{Neutrino Energy Loss in Stellar Interiors. VII. Pair,
  Photo-, Plasma, Bremsstrahlung, and Recombination Neutrino Processes},}
  \apjs, 102, 411, \dodoi{10.1086/192264}

% type= phdthesis
\bibitem[{N. {Ivanova}(2002){Ivanova}}]{2002PhDT........25I}
{Ivanova}, N. 2002, \bibinfo{title}{{Slow mergers of massive stars},} PhD
  thesis, University of Oxford, UK

% type= article
\bibitem[{N. Ivanova(2011)Ivanova}]{ivanova_core}
Ivanova, N. 2011, \bibinfo{title}{Common envelope: on the mass and the fate of
  the remnant,} \apj, 730, 76, \dodoi{10.1088/0004-637X/730/2/76}

% type= article
\bibitem[{N. Ivanova \& S. Chaichenets(2011)Ivanova \&
  Chaichenets}]{ivanova_enthalpy_2011}
Ivanova, N., \& Chaichenets, S. 2011, \bibinfo{title}{{COMMON} {ENVELOPE}:
  {ENTHALPY} {CONSIDERATION},} \apjl, 731, L36,
  \dodoi{10.1088/2041-8205/731/2/l36}

% type= article
\bibitem[{N. {Ivanova} {et~al.}(2013){Ivanova}, {Justham}, {Avendano Nandez},
  \& {Lombardi}}]{Ivanova2013}
{Ivanova}, N., {Justham}, S., {Avendano Nandez}, J.~L., \& {Lombardi}, J.~C.
  2013, \bibinfo{title}{{Identification of the Long-Sought Common-Envelope
  Events},} Science, 339, 433, \dodoi{10.1126/science.1225540}

% type= article
\bibitem[{N. {Ivanova} {et~al.}(2024){Ivanova}, {Kundu}, \&
  {Pourmand}}]{2024ApJ...971...64I}
{Ivanova}, N., {Kundu}, S., \& {Pourmand}, A. 2024, \bibinfo{title}{{Unified
  Rapid Mass Transfer},} \apj, 971, 64, \dodoi{10.3847/1538-4357/ad583e}

% type= article
\bibitem[{N. {Ivanova} \& J.~L.~A. {Nandez}(2016){Ivanova} \&
  {Nandez}}]{2016MNRAS.462..362I}
{Ivanova}, N., \& {Nandez}, J.~L.~A. 2016, \bibinfo{title}{{Common envelope
  events with low-mass giants: understanding the transition to the slow
  spiral-in},} \mnras, 462, 362, \dodoi{10.1093/mnras/stw1676}

% type= inproceedings
\bibitem[{N. {Ivanova} \& P. {Podsiadlowski}(2003){Ivanova} \&
  {Podsiadlowski}}]{Ivanova2003}
{Ivanova}, N., \& {Podsiadlowski}, P. 2003, \bibinfo{title}{{The Slow Merger of
  Massive Stars},} in From Twilight to Highlight: The Physics of Supernovae,
  ed. W.~{Hillebrandt} \& B.~{Leibundgut}, 19, \dodoi{10.1007/10828549_3}

% type= article
\bibitem[{N. {Ivanova} {et~al.}(2002){Ivanova}, {Podsiadlowski}, \&
  {Spruit}}]{2002MNRAS.334..819I}
{Ivanova}, N., {Podsiadlowski}, P., \& {Spruit}, H. 2002,
  \bibinfo{title}{{Hydrodynamical simulations of the stream-core interaction in
  the slow merger of massive stars},} \mnras, 334, 819,
  \dodoi{10.1046/j.1365-8711.2002.05543.x}

% type= article
\bibitem[{A.~S. {Jermyn} {et~al.}(2021){Jermyn}, {Schwab}, {Bauer}, {Timmes},
  \& {Potekhin}}]{Jermyn2021}
{Jermyn}, A.~S., {Schwab}, J., {Bauer}, E., {Timmes}, F.~X., \& {Potekhin},
  A.~Y. 2021, \bibinfo{title}{{Skye: A Differentiable Equation of State},}
  \apj, 913, 72, \dodoi{10.3847/1538-4357/abf48e}

% type= article
\bibitem[{A.~S. {Jermyn} {et~al.}(2023){Jermyn}, {Bauer}, {Schwab}, {Farmer},
  {Ball}, {Bellinger}, {Dotter}, {Joyce}, {Marchant}, {Mombarg}, {Wolf}, {Sunny
  Wong}, {Cinquegrana}, {Farrell}, {Smolec}, {Thoul}, {Cantiello}, {Herwig},
  {Toloza}, {Bildsten}, {Townsend}, \& {Timmes}}]{Jermyn2023}
{Jermyn}, A.~S., {Bauer}, E.~B., {Schwab}, J., {et~al.} 2023,
  \bibinfo{title}{{Modules for Experiments in Stellar Astrophysics (MESA):
  Time-dependent Convection, Energy Conservation, Automatic Differentiation,
  and Infrastructure},} \apjs, 265, 15, \dodoi{10.3847/1538-4365/acae8d}

% type= article
\bibitem[{D. {Jones} {et~al.}(2015){Jones}, {Boffin}, {Rodr{\'\i}guez-Gil},
  {Wesson}, {Corradi}, {Miszalski}, \& {Mohamed}}]{2015A&A...580A..19J}
{Jones}, D., {Boffin}, H.~M.~J., {Rodr{\'\i}guez-Gil}, P., {et~al.} 2015,
  \bibinfo{title}{{The post-common envelope central stars of the planetary
  nebulae Henize 2-155 and Henize 2-161},} \aap, 580, A19,
  \dodoi{10.1051/0004-6361/201425454}

% type= article
\bibitem[{R. {Kippenhahn} \& A. {Weigert}(1967){Kippenhahn} \&
  {Weigert}}]{1967ZA.....65..251K}
{Kippenhahn}, R., \& {Weigert}, A. 1967, \bibinfo{title}{{Entwicklung in engen
  Doppelsternsystemen I. Massenaustausch vor und nach Beendigung des zentralen
  Wasserstoff-Brennens},} \zap, 65, 251

% type= article
\bibitem[{K. {Langanke} \& G. {Mart{\'{\i}}nez-Pinedo}(2000){Langanke} \&
  {Mart{\'{\i}}nez-Pinedo}}]{Langanke2000}
{Langanke}, K., \& {Mart{\'{\i}}nez-Pinedo}, G. 2000,
  \bibinfo{title}{{Shell-model calculations of stellar weak interaction rates:
  II. Weak rates for nuclei in the mass range /A=45-65 in supernovae
  environments},} Nuclear Physics A, 673, 481,
  \dodoi{10.1016/S0375-9474(00)00131-7}

% type= article
\bibitem[{N. {Langer}(1998){Langer}}]{Langer1998}
{Langer}, N. 1998, \bibinfo{title}{{Coupled mass and angular momentum loss of
  massive main sequence stars},} \aap, 329, 551

% type= article
\bibitem[{Z. {Li} {et~al.}(2024){Li}, {Zhu}, {L{\"u}}, {Li}, {Liu}, {Guo},
  {Yu}, \& {Lu}}]{Li2024}
{Li}, Z., {Zhu}, C., {L{\"u}}, G., {et~al.} 2024, \bibinfo{title}{{The
  Population Synthesis of Wolf{\textendash}Rayet Stars Involving Binary Merger
  Channels},} \apj, 969, 160, \dodoi{10.3847/1538-4357/ad4da8}

% type= article
\bibitem[{N. {Lohev} {et~al.}(2019){Lohev}, {Sabach}, {Gilkis}, \&
  {Soker}}]{2019MNRAS.490....9L}
{Lohev}, N., {Sabach}, E., {Gilkis}, A., \& {Soker}, N. 2019,
  \bibinfo{title}{{Type IIb supernova progenitors by fatal common envelope
  evolution},} \mnras, 490, 9, \dodoi{10.1093/mnras/stz2593}

% type= article
\bibitem[{J.~C. {Lombardi} {et~al.}(1996){Lombardi}, {Rasio}, \&
  {Shapiro}}]{1996ApJ...468..797L}
{Lombardi}, Jr., J.~C., {Rasio}, F.~A., \& {Shapiro}, S.~L. 1996,
  \bibinfo{title}{{Collisions of Main-Sequence Stars and the Formation of Blue
  Stragglers in Globular Clusters},} \apj, 468, 797, \dodoi{10.1086/177736}

% type= article
\bibitem[{J.~C. {Lombardi} {et~al.}(2002){Lombardi}, {Warren}, {Rasio},
  {Sills}, \& {Warren}}]{2002ApJ...568..939L}
{Lombardi}, Jr., J.~C., {Warren}, J.~S., {Rasio}, F.~A., {Sills}, A., \&
  {Warren}, A.~R. 2002, \bibinfo{title}{{Stellar Collisions and the Interior
  Structure of Blue Stragglers},} \apj, 568, 939, \dodoi{10.1086/339060}

% type= article
\bibitem[{W. {Lu} {et~al.}(2023){Lu}, {Fuller}, {Quataert}, \&
  {Bonnerot}}]{Lu2023}
{Lu}, W., {Fuller}, J., {Quataert}, E., \& {Bonnerot}, C. 2023,
  \bibinfo{title}{{On rapid binary mass transfer - I. Physical model},} \mnras,
  519, 1409, \dodoi{10.1093/mnras/stac3621}

% type= article
\bibitem[{A. {Maeder} \& G. {Meynet}(2000){Maeder} \&
  {Meynet}}]{MaederMeynet2000}
{Maeder}, A., \& {Meynet}, G. 2000, \bibinfo{title}{{Stellar evolution with
  rotation. VI. The Eddington and Omega -limits, the rotational mass loss for
  OB and LBV stars},} \aap, 361, 159, \dodoi{10.48550/arXiv.astro-ph/0006405}

% type= article
\bibitem[{P. {Massey} {et~al.}(2023){Massey}, {Neugent}, \&
  {Morrell}}]{2023ApJ...947...77M}
{Massey}, P., {Neugent}, K.~F., \& {Morrell}, N.~I. 2023,
  \bibinfo{title}{{Constraints on the Binarity of the WN3/O3 Class of
  Wolf-Rayet Stars},} \apj, 947, 77, \dodoi{10.3847/1538-4357/acc5ed}

% type= article
\bibitem[{A. {McKeown} {et~al.}(2026){McKeown}, {Furey}, {Sabhahit}, \&
  {Vink}}]{2026A&A...715L...1M}
{McKeown}, A., {Furey}, C., {Sabhahit}, G.~N., \& {Vink}, J.~S. 2026,
  \bibinfo{title}{{Wolf-Rayet formation through L/M-dependent superwinds},}
  \aap, 715, L1, \dodoi{10.1051/0004-6361/202661554}

% type= article
\bibitem[{A. {Menon} \& A. {Heger}(2017){Menon} \&
  {Heger}}]{2017MNRAS.469.4649M}
{Menon}, A., \& {Heger}, A. 2017, \bibinfo{title}{{The quest for blue
  supergiants: binary merger models for the evolution of the progenitor of SN
  1987A.},} \mnras, 469, 4649, \dodoi{10.1093/mnras/stx818}

% type= article
\bibitem[{P.~E. {M{\"u}ller} \& J.~S. {Vink}(2014){M{\"u}ller} \&
  {Vink}}]{MullerVink2014}
{M{\"u}ller}, P.~E., \& {Vink}, J.~S. 2014, \bibinfo{title}{{Rotating massive O
  stars with non-spherical 2D winds},} \aap, 564, A57,
  \dodoi{10.1051/0004-6361/201323031}

% type= article
\bibitem[{K.~F. {Neugent} {et~al.}(2017){Neugent}, {Massey}, {Hillier}, \&
  {Morrell}}]{Neugent2017}
{Neugent}, K.~F., {Massey}, P., {Hillier}, D.~J., \& {Morrell}, N. 2017,
  \bibinfo{title}{{The Evolution and Physical Parameters of WN3/O3s: A New Type
  of Wolf-Rayet Star},} \apj, 841, 20, \dodoi{10.3847/1538-4357/aa6e51}

% type= article
\bibitem[{K.~I. {Nomoto} {et~al.}(1995){Nomoto}, {Iwamoto}, \&
  {Suzuki}}]{Nomoto1995}
{Nomoto}, K.~I., {Iwamoto}, K., \& {Suzuki}, T. 1995, \bibinfo{title}{{The
  evolution and explosion of massive binary stars and Type Ib-Ic-IIb-IIL
  supernovae.},} \physrep, 256, 173, \dodoi{10.1016/0370-1573(94)00107-E}

% type= article
\bibitem[{T. {Nugis} \& H.~J.~G.~L.~M. {Lamers}(2000){Nugis} \&
  {Lamers}}]{Nugis2000}
{Nugis}, T., \& {Lamers}, H.~J.~G.~L.~M. 2000, \bibinfo{title}{{Mass-loss rates
  of Wolf-Rayet stars as a function of stellar parameters},} \aap, 360, 227

% type= article
\bibitem[{T. {Oda} {et~al.}(1994){Oda}, {Hino}, {Muto}, {Takahara}, \&
  {Sato}}]{Oda1994}
{Oda}, T., {Hino}, M., {Muto}, K., {Takahara}, M., \& {Sato}, K. 1994,
  \bibinfo{title}{{Rate Tables for the Weak Processes of sd-Shell Nuclei in
  Stellar Matter},} Atomic Data and Nuclear Data Tables, 56, 231,
  \dodoi{10.1006/adnd.1994.1007}

% type= article
\bibitem[{S.~P. {Owocki} {et~al.}(1996){Owocki}, {Cranmer}, \&
  {Gayley}}]{Owocki1996}
{Owocki}, S.~P., {Cranmer}, S.~R., \& {Gayley}, K.~G. 1996,
  \bibinfo{title}{{Inhibition FO Wind Compressed Disk Formation by Nonradial
  Line-Forces in Rotating Hot-Star Winds},} \apjl, 472, L115,
  \dodoi{10.1086/310372}

% type= article
\bibitem[{B. {Paczy{\'n}ski}(1965){Paczy{\'n}ski}}]{Pacz1965}
{Paczy{\'n}ski}, B. 1965, \bibinfo{title}{{Cataclysmic Variables among Binary
  Stars I. U Geminorum Stars},} \actaa, 15, 89

% type= article
\bibitem[{B. {Paczy{\'n}ski}(1967){Paczy{\'n}ski}}]{1967AcA....17..355P}
{Paczy{\'n}ski}, B. 1967, \bibinfo{title}{{Evolution of Close Binaries. V. The
  Evolution of Massive Binaries and the Formation of the Wolf-Rayet Stars},}
  \actaa, 17, 355

% type= inproceedings
\bibitem[{B. Paczynski(1976)Paczynski}]{paczynski_common_1976}
Paczynski, B. 1976, \bibinfo{title}{Common {Envelope} {Binaries},} in {IAU}
  {Symposium}, Vol.~73, Structure and {Evolution} of {Close} {Binary}
  {Systems}, ed. P.~Eggleton, S.~Mitton, \& J.~Whelan, 75

% type= article
\bibitem[{D. Pauli {et~al.}(2022)Pauli, Langer, Aguilera-Dena, Wang, \&
  Marchant}]{Pauli_2022}
Pauli, D., Langer, N., Aguilera-Dena, D.~R., Wang, C., \& Marchant, P. 2022,
  \bibinfo{title}{A synthetic population of Wolf-Rayet stars in the LMC based
  on detailed single and binary star evolution models,} A\&A, 667, A58,
  \dodoi{10.1051/0004-6361/202243965}

% type= article
\bibitem[{D. {Pauli} {et~al.}(2026){Pauli}, {Langer}, {Schootemeijer},
  {Marchant}, {Jin}, {Ercolino}, {Picco}, {Willcox}, \& {Sana}}]{Pauli2026}
{Pauli}, D., {Langer}, N., {Schootemeijer}, A., {et~al.} 2026,
  \bibinfo{title}{{The drastic impact of Eddington-limit induced mass ejections
  on massive star populations},} \aap, 707, A11,
  \dodoi{10.1051/0004-6361/202557766}

% type= article
\bibitem[{B. {Paxton} {et~al.}(2011){Paxton}, {Bildsten}, {Dotter}, {Herwig},
  {Lesaffre}, \& {Timmes}}]{Paxton2011}
{Paxton}, B., {Bildsten}, L., {Dotter}, A., {et~al.} 2011,
  \bibinfo{title}{{Modules for Experiments in Stellar Astrophysics (MESA)},}
  \apjs, 192, 3, \dodoi{10.1088/0067-0049/192/1/3}

% type= article
\bibitem[{B. {Paxton} {et~al.}(2013){Paxton}, {Cantiello}, {Arras}, {Bildsten},
  {Brown}, {Dotter}, {Mankovich}, {Montgomery}, {Stello}, {Timmes}, \&
  {Townsend}}]{Paxton2013}
{Paxton}, B., {Cantiello}, M., {Arras}, P., {et~al.} 2013,
  \bibinfo{title}{{Modules for Experiments in Stellar Astrophysics (MESA):
  Planets, Oscillations, Rotation, and Massive Stars},} \apjs, 208, 4,
  \dodoi{10.1088/0067-0049/208/1/4}

% type= article
\bibitem[{B. {Paxton} {et~al.}(2015){Paxton}, {Marchant}, {Schwab}, {Bauer},
  {Bildsten}, {Cantiello}, {Dessart}, {Farmer}, {Hu}, {Langer}, {Townsend},
  {Townsley}, \& {Timmes}}]{Paxton2015}
{Paxton}, B., {Marchant}, P., {Schwab}, J., {et~al.} 2015,
  \bibinfo{title}{{Modules for Experiments in Stellar Astrophysics (MESA):
  Binaries, Pulsations, and Explosions},} \apjs, 220, 15,
  \dodoi{10.1088/0067-0049/220/1/15}

% type= article
\bibitem[{B. {Paxton} {et~al.}(2018){Paxton}, {Schwab}, {Bauer}, {Bildsten},
  {Blinnikov}, {Duffell}, {Farmer}, {Goldberg}, {Marchant}, {Sorokina},
  {Thoul}, {Townsend}, \& {Timmes}}]{Paxton2018}
{Paxton}, B., {Schwab}, J., {Bauer}, E.~B., {et~al.} 2018,
  \bibinfo{title}{{Modules for Experiments in Stellar Astrophysics (MESA):
  Convective Boundaries, Element Diffusion, and Massive Star Explosions},}
  \apjs, 234, 34, \dodoi{10.3847/1538-4365/aaa5a8}

% type= article
\bibitem[{B. {Paxton} {et~al.}(2019){Paxton}, {Smolec}, {Schwab}, {Gautschy},
  {Bildsten}, {Cantiello}, {Dotter}, {Farmer}, {Goldberg}, {Jermyn}, {Kanbur},
  {Marchant}, {Thoul}, {Townsend}, {Wolf}, {Zhang}, \& {Timmes}}]{Paxton2019}
{Paxton}, B., {Smolec}, R., {Schwab}, J., {et~al.} 2019,
  \bibinfo{title}{{Modules for Experiments in Stellar Astrophysics (MESA):
  Pulsating Variable Stars, Rotation, Convective Boundaries, and Energy
  Conservation},} \apjs, 243, 10, \dodoi{10.3847/1538-4365/ab2241}

% type= article
\bibitem[{P. {Podsiadlowski} {et~al.}(1990){Podsiadlowski}, {Joss}, \&
  {Rappaport}}]{1990A&A...227L...9P}
{Podsiadlowski}, P., {Joss}, P.~C., \& {Rappaport}, S. 1990, \bibinfo{title}{{A
  merger model for SN 1987A.},} \aap, 227, L9

% type= article
\bibitem[{A.~Y. {Potekhin} \& G. {Chabrier}(2010){Potekhin} \&
  {Chabrier}}]{Potekhin2010}
{Potekhin}, A.~Y., \& {Chabrier}, G. 2010, \bibinfo{title}{{Thermodynamic
  Functions of Dense Plasmas: Analytic Approximations for Astrophysical
  Applications},} Contributions to Plasma Physics, 50, 82,
  \dodoi{10.1002/ctpp.201010017}

% type= article
\bibitem[{A. {Pourmand} \& N. {Ivanova}(2023){Pourmand} \&
  {Ivanova}}]{2023ApJ...952..126P}
{Pourmand}, A., \& {Ivanova}, N. 2023, \bibinfo{title}{{Properties of Binary
  Systems in a One-dimensional Approximation},} \apj, 952, 126,
  \dodoi{10.3847/1538-4357/acd4c1}

% type= article
\bibitem[{J. {Poutanen}(2017){Poutanen}}]{Poutanen2017}
{Poutanen}, J. 2017, \bibinfo{title}{{Rosseland and Flux Mean Opacities for
  Compton Scattering},} \apj, 835, 119, \dodoi{10.3847/1538-4357/835/2/119}

% type= incollection
\bibitem[{D. {Reimers}(1975){Reimers}}]{Reimers1975}
{Reimers}, D. 1975, \bibinfo{title}{{Circumstellar envelopes and mass loss of
  red giant stars.},} in Problems in stellar atmospheres and envelopes., ed.
  B.~{Baschek}, W.~H. {Kegel}, \& G.~{Traving}, 229--256

% type= article
\bibitem[{F.~J. {Rogers} \& A. {Nayfonov}(2002){Rogers} \&
  {Nayfonov}}]{Rogers2002}
{Rogers}, F.~J., \& {Nayfonov}, A. 2002, \bibinfo{title}{{Updated and Expanded
  OPAL Equation-of-State Tables: Implications for Helioseismology},} \apj, 576,
  1064, \dodoi{10.1086/341894}

% type= article
\bibitem[{G.~N. {Sabhahit} {et~al.}(2026){Sabhahit}, {Vink}, \&
  {Sander}}]{Sabhahit2026}
{Sabhahit}, G.~N., {Vink}, J.~S., \& {Sander}, A. A.~C. 2026,
  \bibinfo{title}{{Hydrodynamical mass-loss rates for Very Massive Stars II.
  New theoretical mass-loss predictions at solar metallicity (Z = 0.02)},}
  arXiv e-prints, arXiv:2607.28012, \dodoi{10.48550/arXiv.2607.28012}

% type= article
\bibitem[{G.~N. {Sabhahit} {et~al.}(2025){Sabhahit}, {Vink}, {Sander}, \&
  {Ramachandran}}]{2025A&A...703A..40S}
{Sabhahit}, G.~N., {Vink}, J.~S., {Sander}, A. A.~C., \& {Ramachandran}, V.
  2025, \bibinfo{title}{{Minimum and maximum mass-luminosity relations for
  stripped stars},} \aap, 703, A40, \dodoi{10.1051/0004-6361/202554852}

% type= article
\bibitem[{A. {Sander} {et~al.}(2015){Sander}, {Shenar}, {Hainich},
  {G{\'\i}menez-Garc{\'\i}a}, {Todt}, \& {Hamann}}]{Sander2015}
{Sander}, A., {Shenar}, T., {Hainich}, R., {et~al.} 2015, \bibinfo{title}{{On
  the consistent treatment of the quasi-hydrostatic layers in hot star
  atmospheres},} \aap, 577, A13, \dodoi{10.1051/0004-6361/201425356}

% type= article
\bibitem[{A.~A.~C. {Sander} \& J.~S. {Vink}(2020){Sander} \&
  {Vink}}]{Sander2020}
{Sander}, A. A.~C., \& {Vink}, J.~S. 2020, \bibinfo{title}{{On the nature of
  massive helium star winds and Wolf-Rayet-type mass-loss},} \mnras, 499, 873,
  \dodoi{10.1093/mnras/staa2712}

% type= article
\bibitem[{D. {Saumon} {et~al.}(1995){Saumon}, {Chabrier}, \& {van
  Horn}}]{Saumon1995}
{Saumon}, D., {Chabrier}, G., \& {van Horn}, H.~M. 1995, \bibinfo{title}{{An
  Equation of State for Low-Mass Stars and Giant Planets},} \apjs, 99, 713,
  \dodoi{10.1086/192204}

% type= article
\bibitem[{F.~R.~N. {Schneider}(2026){Schneider}}]{2026ARA&A..64..629S}
{Schneider}, F. R.~N. 2026, \bibinfo{title}{{Theory, Simulations, and
  Observations of Stellar Mergers},} \araa, 64, 629,
  \dodoi{10.1146/annurev-astro-051024-050338}

% type= article
\bibitem[{A. Schootemeijer \& N. Langer(2018)Schootemeijer \&
  Langer}]{Schootemeijer_2018}
Schootemeijer, A., \& Langer, N. 2018, \bibinfo{title}{Wolf–Rayet stars in
  the Small Magellanic Cloud as testbed for massive star evolution,} A\&A, 611,
  A75, \dodoi{10.1051/0004-6361/201731895}

% type= article
\bibitem[{A. {Schootemeijer} {et~al.}(2024){Schootemeijer}, {Shenar}, {Langer},
  {Grin}, {Sana}, {Gr{\"a}fener}, {Sch{\"u}rmann}, {Wang}, \&
  {Xu}}]{Schootemeijer_2024}
{Schootemeijer}, A., {Shenar}, T., {Langer}, N., {et~al.} 2024,
  \bibinfo{title}{{An absence of binary companions to Wolf-Rayet stars in the
  Small Magellanic Cloud: Implications for mass loss and black hole masses at
  low metallicities},} \aap, 689, A157, \dodoi{10.1051/0004-6361/202449978}

% type= article
\bibitem[{T. Shenar {et~al.}(2020)Shenar, Gilkis, Vink, Sana, \&
  Sander}]{Shenar_2020}
Shenar, T., Gilkis, A., Vink, J.~S., Sana, H., \& Sander, A. A.~C. 2020,
  \bibinfo{title}{Why binary interaction does not necessarily dominate the
  formation of Wolf-Rayet stars at low metallicity,} A\&A, 634, A79,
  \dodoi{10.1051/0004-6361/201936948}

% type= article
\bibitem[{T. {Shenar} {et~al.}(2016){Shenar}, {Hainich}, {Todt}, {Sander},
  {Hamann}, {Moffat}, {Eldridge}, {Pablo}, {Oskinova}, \&
  {Richardson}}]{2016A&A...591A..22S}
{Shenar}, T., {Hainich}, R., {Todt}, H., {et~al.} 2016,
  \bibinfo{title}{{Wolf-Rayet stars in the Small Magellanic Cloud. II. Analysis
  of the binaries},} \aap, 591, A22, \dodoi{10.1051/0004-6361/201527916}

% type= article
\bibitem[{T. {Shenar} {et~al.}(2019){Shenar}, {Sablowski}, {Hainich}, {Todt},
  {Moffat}, {Oskinova}, {Ramachandran}, {Sana}, {Sander}, {Schnurr},
  {St-Louis}, {Vanbeveren}, {G{\"o}tberg}, \& {Hamann}}]{2019A&A...627A.151S}
{Shenar}, T., {Sablowski}, D.~P., {Hainich}, R., {et~al.} 2019,
  \bibinfo{title}{{The Wolf-Rayet binaries of the nitrogen sequence in the
  Large Magellanic Cloud. Spectroscopy, orbital analysis, formation, and
  evolution},} \aap, 627, A151, \dodoi{10.1051/0004-6361/201935684}

% type= article
\bibitem[{T. {Shenar} {et~al.}(2020){Shenar}, {Sablowski}, {Hainich}, {Todt},
  {Moffat}, {Oskinova}, {Ramachandran}, {Sana}, {Sander}, {Schnurr},
  {St-Louis}, {Vanbeveren}, {G{\"o}tberg}, \& {Hamann}}]{Shenar2020_Erratum}
{Shenar}, T., {Sablowski}, D.~P., {Hainich}, R., {et~al.} 2020,
  \bibinfo{title}{{The Wolf-Rayet binaries of the nitrogen sequence in the
  Large Magellanic Cloud. Spectroscopy, orbital analysis, formation, and
  evolution (Corrigendum)},} \aap, 641, C2,
  \dodoi{10.1051/0004-6361/201935684e}

% type= article
\bibitem[{T. {Shenar} {et~al.}(2023){Shenar}, {Wade}, {Marchant}, {Bagnulo},
  {Bodensteiner}, {Bowman}, {Gilkis}, {Langer}, {Nicolas-Chen{\'e}},
  {Oskinova}, {Van Reeth}, {Sana}, {St-Louis}, {de Oliveira}, {Todt}, \&
  {Toonen}}]{Shenar2023}
{Shenar}, T., {Wade}, G.~A., {Marchant}, P., {et~al.} 2023, \bibinfo{title}{{A
  massive helium star with a sufficiently strong magnetic field to form a
  magnetar},} Science, 381, 761, \dodoi{10.1126/science.ade3293}

% type= article
\bibitem[{N. {Smith} {et~al.}(2018){Smith}, {G{\"o}tberg}, \& {de
  Mink}}]{Smith_2017}
{Smith}, N., {G{\"o}tberg}, Y., \& {de Mink}, S.~E. 2018,
  \bibinfo{title}{{Extreme isolation of WN3/O3 stars and implications for their
  evolutionary origin as the elusive stripped binaries},} \mnras, 475, 772,
  \dodoi{10.1093/mnras/stx3181}

% type= article
\bibitem[{N. {Smith} {et~al.}(2016){Smith}, {Andrews}, {Van Dyk}, {Mauerhan},
  {Kasliwal}, {Bond}, {Filippenko}, {Clubb}, {Graham}, {Perley}, {Jencson},
  {Bally}, {Ubeda}, \& {Sabbi}}]{Smith2016}
{Smith}, N., {Andrews}, J.~E., {Van Dyk}, S.~D., {et~al.} 2016,
  \bibinfo{title}{{Massive star mergers and the recent transient in NGC 4490: a
  more massive cousin of V838 Mon and V1309 Sco},} \mnras, 458, 950,
  \dodoi{10.1093/mnras/stw219}

% type= article
\bibitem[{N. {Soker}(2019){Soker}}]{2019SCPMA..6219501S}
{Soker}, N. 2019, \bibinfo{title}{{The class of supernova progenitors that
  result from fatal common envelope evolution},} Science China Physics,
  Mechanics, and Astronomy, 62, 119501, \dodoi{10.1007/s11433-019-9402-x}

% type= article
\bibitem[{F.~X. {Timmes} \& F.~D. {Swesty}(2000){Timmes} \&
  {Swesty}}]{Timmes2000}
{Timmes}, F.~X., \& {Swesty}, F.~D. 2000, \bibinfo{title}{{The Accuracy,
  Consistency, and Speed of an Electron-Positron Equation of State Based on
  Table Interpolation of the Helmholtz Free Energy},} \apjs, 126, 501,
  \dodoi{10.1086/313304}

% type= article
\bibitem[{F. {Tramper} {et~al.}(2015){Tramper}, {Straal}, {Sanyal}, {Sana}, {de
  Koter}, {Gr{\"a}fener}, {Langer}, {Vink}, {de Mink}, \&
  {Kaper}}]{2015AA...581A.110T}
{Tramper}, F., {Straal}, S.~M., {Sanyal}, D., {et~al.} 2015,
  \bibinfo{title}{{Massive stars on the verge of exploding: the properties of
  oxygen sequence Wolf-Rayet stars},} \aap, 581, A110,
  \dodoi{10.1051/0004-6361/201425390}

% type= article
\bibitem[{R. {Trigg} {et~al.}(2026){Trigg}, {Oskinova}, {Hamann}, {Todt},
  {Pauli}, {Shenar}, {Sander}, {Chatzis}, {Mang}, {Lefever}, \&
  et~al.}]{2026arXiv260922918T}
{Trigg}, R., {Oskinova}, L.~M., {Hamann}, W.-R., {et~al.} 2026,
  \bibinfo{title}{{Distinct Population of Wolf-Rayet Stars in the
  Low-Metallicity Galaxy NGC 6822. Quantitative analysis of optical and UV
  spectra of the complete sample},} arXiv e-prints, arXiv:2609.22918.
\newblock \doarXiv{2609.22918}

% type= article
\bibitem[{R. {Tylenda} {et~al.}(2011){Tylenda}, {Hajduk}, {Kami{\'n}ski},
  {Udalski}, {Soszy{\'n}ski}, {Szyma{\'n}ski}, {Kubiak}, {Pietrzy{\'n}ski},
  {Poleski}, {Wyrzykowski}, \& {Ulaczyk}}]{Tylenda2011}
{Tylenda}, R., {Hajduk}, M., {Kami{\'n}ski}, T., {et~al.} 2011,
  \bibinfo{title}{{V1309 Scorpii: merger of a contact binary},} \aap, 528,
  A114, \dodoi{10.1051/0004-6361/201016221}

% type= article
\bibitem[{J.~S. {Vink}(2017){Vink}}]{Vink2017}
{Vink}, J.~S. 2017, \bibinfo{title}{{Winds from stripped low-mass helium stars
  and Wolf-Rayet stars},} \aap, 607, L8, \dodoi{10.1051/0004-6361/201731902}

% type= article
\bibitem[{J.~S. {Vink} {et~al.}(2001){Vink}, {de Koter}, \&
  {Lamers}}]{Vink2001}
{Vink}, J.~S., {de Koter}, A., \& {Lamers}, H.~J.~G.~L.~M. 2001,
  \bibinfo{title}{{Mass-loss predictions for O and B stars as a function of
  metallicity},} \aap, 369, 574, \dodoi{10.1051/0004-6361:20010127}

% type= article
\bibitem[{R.~F. {Webbink}(1984){Webbink}}]{Webbink84}
{Webbink}, R.~F. 1984, \bibinfo{title}{{Double white dwarfs as progenitors of R
  Coronae Borealis stars and Type I supernovae},} \apj, 277, 355,
  \dodoi{10.1086/161701}

% type= article
\bibitem[{C.~J.~E. {Wolf} \& G. {Rayet}(1867){Wolf} \&
  {Rayet}}]{1867CRAS...65..292W}
{Wolf}, C.~J.~E., \& {Rayet}, G. 1867, \bibinfo{title}{{Spectroscopie
  stellaire},} Academie des Sciences Paris Comptes Rendus, 65, 292

\end{thebibliography}

\end{document}